\documentclass[graybox]{svmult}

\usepackage{type1cm}        
\usepackage{makeidx}         
\usepackage{graphicx}        
\usepackage{multicol}        
\usepackage[bottom]{footmisc}

\usepackage{newtxtext}       %
\usepackage{newtxmath}       

\makeindex             

\begin{document}

\title*{Action Research with Industrial Software Engineering 
--- An Educational Perspective}
\author{Yvonne Dittrich, Johan Bolmsten and Cathrine Seidelin}
\institute{Yvonne  Dittrich \at IT University of Copenhagen, Rued Langgaards Vej 7, 2300 Copenhagen S, Denmark \email{ydi@itu.dk}
\and Johan Bolmsten \at World Maritime University. Fiskehamnsgatan 1, 211 18 Malmö \email{jb@wmu.se} \and
Cathrine Seidelin \at Novo Nordisk, Novo Alle 1, 2880 Bagsværd, Denmark  \email{cathrineseidelin@gmail.com}}
%
%
\maketitle

\abstract*{Action research provides the opportunity to explore the usefulness and usability of software engineering methods in industrial settings, and makes it possible to develop methods, tools and techniques with software engineering practitioners. However, as the research moves beyond the observational approach, it requires a different kind of interaction with the software development organisation. This makes action research a challenging endeavour, and it makes it difficult to teach action research through a course that goes beyond explaining the principles.
\\
This chapter is intended to support learning and teaching action research, by providing a rich set of examples, and identifying tools that we found helpful in our action research projects. The core of this chapter focusses on our interaction with the participating developers and domain experts, and the organisational setting.
\\
This chapter is structured around a set of challenges that reoccurred in the action research projects in which the authors participated. Each section is accompanied by a toolkit that presents related techniques and tools. The exercises are designed to explore the topics, and practise using the tools and techniques presented. We hope the material in this chapter encourages researchers who are new to action research to further explore this promising opportunity. }

\abstract{Action research provides the opportunity to explore the usefulness and usability of software engineering methods in industrial settings, and makes it possible to develop methods, tools and techniques with software engineering practitioners. However, as the research moves beyond the observational approach, it requires a different kind of interaction with the software development organisation. This makes action research a challenging endeavour, and it makes it difficult to teach action research through a course that goes beyond explaining the principles.
\\
This chapter is intended to support learning and teaching action research, by providing a rich set of examples, and identifying tools that we found helpful in our action research projects. The core of this chapter focusses on our interaction with the participating developers and domain experts, and the organisational setting.
\\
This chapter is structured around a set of challenges that reoccurred in the action research projects in which the authors participated. Each section is accompanied by a toolkit that presents related techniques and tools. The exercises are designed to explore the topics, and practise using the tools and techniques presented. We hope the material in this chapter encourages researchers who are new to action research to further explore this promising opportunity. 
\\
\\
A revised version of the chapter is published in 
Daniel Méndez, Paris Avgeriou, Marcos Kalinowski, Nauman Bin Ali:
Handbook on Teaching Empirical Software Engineering. Springer Nature Switzerland 2024, ISBN 978-3-031-71768-0
\\
The publisher's editing change the chapter significantly. Please, refer to the published version for reference and citation.}

\section{Introduction}
\label{sec:1Introduction}
This chapter deepens the understanding of action research in industrial software engineering with rich examples, and emphasises the practicalities of action research in and with industrial software engineering. The chapter mainly builds on three action research projects that are also used as cases that provide examples to illustrate themes highlighted in this chapter. The first project, The SIM Case, addressed software architecture methods that support the evolvability of a software product that implements hydraulic simulations of water systems (rivers, fresh water and sewer systems). It has been the basis of a PhD dissertation \cite{Unphon2010a}  and several articles \cite{UnphonDittrich2010, UphonDittrich2008, Unphon2009a}. The second project, The WMU Case, introduced software engineering and IT management methods of in-house participatory design and user-centred development to an academic capacity-building context \cite{Bolmsten2016, BolmstenDittrich2011, BolmstenDittrich2015}. The third project, The Data CoDesign Case, explored how domain experts may use data and data analytics to support the innovation and development of vocational education and training \cite{Seidelin2020, SeidelinDittrichGronval2020, SeidelinSivertsenDittrich2020}. Yvonne Dittrich was the PhD supervisor in all three cases. The cases are introduced in the first three sections of this chapter. However, all the cases have informed all the sections, and we report episodes from various cases to illustrate the theme of each of the sections.
All three projects implemented, or were inspired by, the Cooperative Method Development (CMD) approach, which first has been introduced in \cite{CMD2008}. Where appropriate, we also use examples from the projects that were the basis of the development of the CMD article.

CMD is an action research approach that combines qualitative empirical research with software engineering tools, methods and process improvements \cite{CMD2008}. CMD is a structured methodological framework that is designed to cope with the complexities of action research in an academically rigorous manner (also see the work by Checkland and Howell \cite{checkland2007action} and by Mathiassen \cite{mathiassen2002collaborative}). Five guidelines define the structure of CMD: (1) an action research cycle consisting of three phases (understanding, deliberating change, implementing and evaluating improvements); (2) ethnographically-inspired research complemented by other methods, if suitable; (3) a focus on shop-floor development practices; (4) deliberating improvement with involved practitioners; (5) assuming the practitioner’s perspective when evaluating the empirical research and deliberating improvements. Figure 1 illustrates the CMD approach: The empirical research results in an understanding of the current practices that inform the joint definition of one or more problems to be addressed by the action research. Based on this problem identification, an intervention or improvement is developed collaboratively by the researcher and practitioners. The researchers contribute with the state of the art research related to the problem identified, based on their survey of the research discourse. The discussions during this deliberation may provide input to the research. Together, practitioners and researchers implement the improvement and evaluate it. This too is informed by research, and it contributes findings and insights that further the research discourse.

\begin{figure}[b]

\centering
\includegraphics[scale=1]{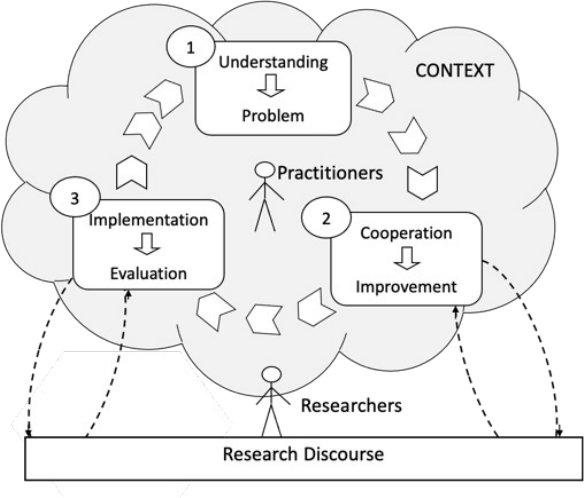}

\label{fig:1CMD}      
\caption{ Illustration of the Cooperative Method Development approach to action research The numbers indicate the 3 phases of the CMD cycle \cite[p. 21]{Eriksson2008}.}
\end{figure}

The Chapter complements Staron’s introductory chapter on action research, which lays the foundation for action research and how to teach it. It is presumed that the reader has studied Staron’s chapter or a similar introduction into action research before studying this one.
It shares the emphasis on Participatory Action Research of Staron’s chapter: one of the core principles of participatory action research is the involvement of the practitioners and domain experts whose work practices the interventions address, in the discussion and decisions about these interventions. 

However, the CMD approach is more explicit about the choice of empirical research to understand the current situation from a participant’s point of view, and evaluate the intervention, again using an ethnographically inspired approach. This may be due to its historical development: When we started to apply qualitative empirical research to software development practices in the late 1990s, the companies and software engineers that collaborated with us asked to recommend ways to improve their practices. The CMD approach is designed to support a member’s perspective, also when it moves beyond ethnographical empirical research \cite{Dittrich2002} , to explore improvements by applying and further developing new methods, processes and tools in an ethically responsible and scientifically accountable way. Similarly, the explicit formulation of this version of participatory action research made our research accountable to the management and the software developers with whom we collaborated.

Teaching and learning action research are best done using an apprenticeship model, where the MSc or PhD student, or the researcher new to action research, regularly discusses her involvement with a software engineering team, or an open-source project with the supervisor, a colleague researcher experienced in action research. This chapter reports our experience of learning and teaching action research, and may be most useful for an experienced action researcher who is beginning to supervise action research projects in software engineering and his or her students. Throughout this chapter, we will add to a toolkit of methods and techniques that helped us to cope with the often-challenging situations we encountered in our research and as supervisors. Thus, the toolkit aims to support learning as it allows to exercise some aspects of action research.

For students and teachers of empirical research courses, this chapter provides a rich set of concrete episodes that supports the discussion of challenging situations in class. Teaching action research through a course presents a challenge for educators, when compared to other empirical research methods, for example, experiments, as it is difficult to practise action research at small scale in the classroom. Therefore, in our courses we ask students to formulate a research proposal for an action research project that is connected to their main research focus, as a way to apply what they hear and read about. Alternatively, the teacher could use the course improvement as an example for the whole class. You will find a number of exercise proposals in the remainder of this chapter. They either ask the students to apply the lessons learned in a section by enhancing their project proposal, or implement some of the tools and techniques in the classroom.

We start with general considerations of ways to anchor action research in the organisation, and address the challenges of not only understanding the work practices of the practitioners whose participating in the research, but also the organisation and the internal and external stakeholders. 
The toolkit provided in Section \ref{sec:2ARinIndustry} discusses the formation of a steering committee, the use of ethnographic field work to understand the work practices of the practitioners involved and explore the challenges from a practitioners’ perspective, stakeholder analysis methods to map the wider organisational context, and the formulation of an informed consent form for the participants. The exercise asks you to apply some of this knowledge to the (fictional) project that is presented as part of Exercise 1, below. Two cases are introduced in Section \ref{sec:2ARinIndustry} and the third case is introduced in Section \ref{sec:3Deliberation}. Section \ref{sec:3Deliberation}, Deliberation – Anchoring Interventions and Improvements in the Organisation, is especially relevant to \emph{Participatory} Action Research. We have added an introduction to how to design, implement and make use of workshops to the toolkit. The exercise is designed to allow you to practise designing and implementing workshops. Section \ref{sec:4Visualisations} discusses the use of visualisations as a potent way to foster discussions. The visualisation section adds to the toolkit the presentation of rich pictures, which we use in all our projects, and pointers to additional visualisation techniques. The exercise invites you to add a visual interpretation of the stakeholder analysis of your project proposal. Section \ref{sec:5Challenges} takes up what may be most difficult part of action research: how to handle situations where things do not work out as planned and discussed with the practitioners. This section adds peer debriefing to the toolkit, and the exercise invites you to experience a debriefing situation. Section \ref{sec:6Sustainability}discusses the how action research interventions can lead to sustainable change in the collaborating organisations. The toolkit presents a framework for analysing and discussing the sustainability of change in the organisation. The exercise invites you to discuss the interventions you plan for your project, and to consider how to improve their organisational sustainability. Section \ref{sec:7Writing} ‘From the Action to the Research Results’ concludes this chapter, and highlights ways to ensure that the action research findings and insights may be supported by a scientifically accountable documentation. It adds ethnographically rich points and thematic analysis to the toolkit. The exercise asks you to explore the kinds of contributions you may expect from the research project you developed through part of the exercises.

\begin{question}{Exercise 1}

\noindent Together, some of the exercises in this chapter result in the (improved) design of an action research project proposal. You could reuse and further develop the research project proposal that you have developed, in parallel with your work with Staron’s chapter, which introduces action research.

If you already plan to do action research, feel free to use your project and the proposal you probably developed as a starting point. If you use other methods in your research, try to find an angle to your research that could be executed as action research. For example, if you work with software architecture notations, a related action research project might involve collaborating with a team, and seeing whether you can jointly improve their architecture practices. Another possibility would be to pretend that improving the course is an action research project, and use that fictional action research project as a basis for the exercises.

If you are not building on an existing research proposal, develop a short motivation for your study (about half a page) and formulate a research question related to action research. Describe the (fictional) company and team you plan to collaborate with, or – if you decide to use the course improvement as an exercise – a (fictional) university whose research methods course or PhD programme you would like to improve.
\end{question}

\section{Action Research in the Software Industry}
\label{sec:2ARinIndustry}
Action research in software engineering projects, for private or public organisations, is potentially rewarding, but challenging. It provides an opportunity to make improvements by contributing knowledge and development capacity that are unavailable in the setting where the action research takes place. It is also an opportunity for the action researcher to learn about real-world operating environments and the usefulness of the technologies being developed. At the same time, change in organisations is difficult, and there are many stakeholders with different and sometimes conflicting priorities that need to be addressed. These stakeholders include other software engineers, users and managers. The subsections further discuss what must be considered when starting an action research project with a software development organisation.

\begin{svgraybox}
\textbf{Case 1: Sim – Evolvable Software Products for Hydraulic Simulation.}

\noindent \textbf{\emph{Background.}}
 The research collaboration with SIM was intended to explore methods, tools and techniques, to ensure the evolvability of software products. SIM develops software that models one-, two- and three-dimensional bodies of water, to predict the effect of, for example, dam construction. The software was further developed to simulate near-real time predictions, to support water management. The researchers collaborated with the team responsible for the product that simulated open one-dimensional water systems, such as rivers and creeks. As part of the re-engineering, the company decided to merge this software with a sister product, which facilitated the simulation of closed one-dimensional water systems, such as sewers.

\noindent \textbf{\emph{Action Research.}}
The action research applied the Cooperative Method Development approach \cite{CMD2008}. The action research introduced lightweight software architecture techniques for high-level design, developed a lightweight Architecture Level Evolvability Assessment for focussed discussions of design decisions with relevant stakeholders, and introduced lightweight architecture compliance techniques that use the built system. The research results emphasise the need to adapt software architecture methods and tools to support the continuous evolution of software products: architecture design and evolution takes place as part of everyday software development; architectural practices need to help the software architect to keep up with the changes to the software and the emerging requirements that may challenge the architecture; evolvability is a quality that should be considered during regular software architecture design discussions \cite{Unphon2010, Unphon2009a, UphonDittrich2008}.

\noindent \textbf{\emph{Anchoring the research in the organisation.}}
This project was part of a larger project for the design of evolvable software products, in which another company was also involved. The contact was established by the university’s senior management. The head of software development was the management representative for the joint project’s steering committee. Locally, the tech lead of the open, one-dimensional simulation product was the main contact point.

The research collaboration was part of a broader development to professionalise software development in an organisation where up to that point, the software had been regarded mainly as a tool for the main business, the consultancy that used the software to model and simulate changes to water systems.

The parallel professionalisation affected the project later: Half-way into the 3-year project, the company decided to reorganise. Prior to the reorganisation, the development of the software products was placed with the departments that used the software for consultancy. The reorganisation positioned the software development of various products in an own department. However, this also resulted in the tech lead and main contact leaving the company.

\end{svgraybox}

\subsection{Anchoring the action research project in the organisation}

Action research projects aim to change how organisations develop or use software. Even if the research addresses only the development of one project, such a change may affect the whole organisation and may have an impact on the organisation’s business and economic outcomes. Therefore, it is important to anchor the action research project in the organisation’s management and development. Management needs to be certain that the research does not upset the organisation and its outcomes, but explores new methods and tools that may be expected to benefit the organisation. The development team that collaborates with the researchers, likewise, needs to feel secure in the fact that they are not expected to implement methods that force them to act against their better judgement.

At first glance, the anchoring in the management may appear to contradict the above-mentioned participatory approach to action research. Therefore, when anchoring the project in the organisation, the commitment to the bottom-up improvement needs to be negotiated and agreed on by all parties. This is not only an ethical concern, it also allows the establishment of a situation where the methodological interventions may be based on professional requirements, rather than on (anticipated) management opinion \cite{Dittrich2002}. 

As the SIM case above indicates, the relation between the researchers and the organisation needs to be maintained for the duration of the project. During the three years that an externally-funded project in Europe lasts, changes to the organisation are to be expected. Similarly, participatory action research often develops its own dynamic, and the final outcome may not be what was anticipated by either the company or the researchers involved. Therefore, the dialogue between researchers and the organisation needs to continue throughout the project. One way to ensure and structure an ongoing dialogue is to establish a steering group that includes representatives from both research and practice (see tool box).

At the beginning of a research collaboration, the main purpose of a steering group is to formulate shared goals and align the expectations of the collaboration. The nature of the steering group will develop over time, as the research project matures and the researcher(s) become more integrated into the organisation. Throughout the research project, the steering group is important for ensuring that the research has the necessary organisational backing. Especially, all interventions decided together with the participating team also need to be discussed and agreed on by the steering committee. Also, if other stakeholders, like e.g. business departments or intended users of the software subject to the collaboration should be involved, the steering committee has to secure the organisation’s necessary support. Towards the end of the research collaboration, the steering group is an important forum for ensuring that shared goals are met, and the knowledge gained is shared.

\eject

\begin{svgraybox}
\textbf{Case 2: IU – Domain Experts Co-Designing Data}

\noindent\textbf{\emph{Background.}}
The research collaboration with Industriens Uddannelser (IU) was intended to explore how domain experts (who were not IT professionals) could participate in the design and innovation of the data and data structures that underpinned the data-based services they used, as well as provided to other stakeholders, in their work practices.

The research project originated in the quest to address the societal challenge of advancing small and medium-sized organisations’ capacity to develop and support ways of innovating and designing services by using data more intelligently. IU develops IT systems that support the maintenance and development of vocational and continuing-education programmes related to the industrial sector.

\noindent\textbf{\emph{Action Research.}}
The researcher was heavily involved in the organisation: present at the organisation at least 3 days a week, participating in many internal meetings and other social activities that extended beyond the scope of the action research interventions. The researcher collaborated with several groups at the organisations (see \cite{Ridleyetal2023} for an appraisal and comparison of the project as a long-term action research project that targets infrastructure systems).

The project applied an action research approach as a ‘meta-practice’ \cite{Hayes2018design} and as a process of critical inquiry. The action research was inspired by Robson and McCartan \cite{RobsonMcCartan}, and followed the widely used representation of a spiral or cycle, where each intervention involves three general stages (1) planning a change, (2) implementing the change and observing what happens following the action(s) and (3) reflecting on the processes and the observed changes to plan for further change and the continuation of the cyclical process (see Figure 2). Three cycles were implemented.

\begin{center}  
\includegraphics[scale=.65]{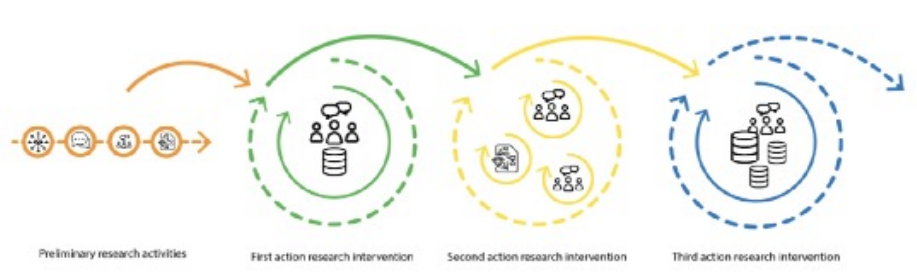}\\
\label{fig:2IU-AR-Cycles}      
{\footnotesize \textbf{Fig. 2} Case 2: Action Research Process \cite[p. 27]{Seidelin2020} }
\end{center}

\noindent\textbf{\emph{Anchoring the research in the organisation.}}
Industriens Uddannelser, contacted the university to obtain support for making better use of various data sources that were available to the organisation. As the organisation was co-owned and financed by labour unions and industrial interest organisations, the management organised a workshop to anchor and scope the research project, even before the application was started.

The first research activities focussed on understanding the organisation’s various administrators’ current ways of working with data. A natural contact for the researcher was a group at the organisation that provided statistics to various committees and working groups. However, other members of the organisation were also involved in the action research cycle.

Throughout the project, the management of Industriens Uddannelser supported the project by allocating resources and time to help the researcher, enabling the co-design of methods with relevant members of the organisation, and by anchoring the project with the external stakeholder.

\end{svgraybox}

\subsection{Understanding the problem before implementing changes}
The first phase of all action research cycles consists of doing empirical research before any intervention takes place. One of its purposes is to be able to document the change brought about by the intervention. If we take the SIM case as an example, to document that the new software architecture and architectural practices are supporting the evolution of the software product in a better way than the existing software architecture and architectural practices, the latter need to be understood and documented. However, the research should not just document a benchmark to compare it with the situation after the intervention. The initial fieldwork, in particular, should be rather broad and take a flexible approach \cite{RobsonMcCartan}, to understand how the software development takes place and the rationale behind the existing practices. We recommend using an ethnographically-inspired approach that emphasises the understanding of a social situation from a member’s point of view: understanding software development from a practitioner’s point of view, and understanding the rationale behind practices that may appear idiosyncratic at first glance will help the researchers to observe and assess both the intended impact of the intervention and the unintended side effects.

One example for the importance to take the practitioner’s perspective seriously is related to the SIM project: the project and the tech lead were reluctant to use architecture documentation. Instead of attributing this to their lack of software engineering training, we conducted an interview study to see what other software product teams do, in terms of architecture documentation \cite{UnphonDittrich2010}. The results helped us to carefully adjust the software architecture practices in order to support communication, rather than replace it \cite{Unphon2009a}.

The initial empirical research is also needed to later customise the intervention for the specific organisation. It provides the basis for understanding what other not-yet-considered factors influence the applicability of specific methods. Referring to the research above: prior to our study, the research on software architecture documentation did emphasise the communication of this architecture to the developers, but did not reflect the effect of a software architecture document on the communication from the software development team to the architect, when they evolved long-living software products. Without prior broad empirical research, we probably would have ignored the team’s reluctance to adopt the use of architecture documentation.

\eject

\begin{tips}{Toolkit Anchoring}
\noindent\textbf{Steering Committee}

\noindent To ensure that all parties involved are regularly informed, and to ensure the necessary resources for a project, the establishment of a steering group for the action research project is crucial. The steering group should meet regularly to discuss the progress of the research and the change initiated, to sanction on the next steps, and to assure the necessary resources in form of time of the development team, access to relevant documents and support for additional research activities.

The steering group should consist of a representative of the management of the collaborating company, a representative of the development, such as the development project’s project manager, with whom the researcher collaborates, the researcher who is implementing the action research, and, if the acting researcher is a PhD student, the supervising professor. The researcher or supervisor would also typically stand for the anchoring of the project with the university. Thus, the steering committee links the actual day-today research with management on both sides, software development organisation and university. The steering committee should be able to take relevant decisions regarding the action research project.

In addition to handling the contractual side of the project with respect to confidentiality, intellectual property and publication procedures, the steering group should also agree on the principal lines of the action research, and, especially, the participatory character of the action research: For the team to freely discuss and adapt methods and tools, they need to be sure that a critical attitude to management’s preferred methods does not adversely affect team members. This also means that the researchers need to negotiate with the organisations management that they do keep the team’s confidentiality and communicate the empirical findings only after they have been cleared with the team.
 \\

\noindent\textbf{Ethnographic fieldwork}

\noindent Ethnographic studies are the subject of this volume’s chapter by Dittrich, Sharp and de Souza. We recommend reading that chapter, and doing the exercises provided there, to plan the ethnographic fieldwork at the beginning of an action research cycle.
\\

\noindent\textbf{Stakeholder analysis}

\noindent Stakeholder analysis is a well-known way to systematically map relevant groups of actors, to consider their interests, for example, when managing a software development project. In strategic management, a number of techniques have been developed to identify stakeholders and analyse their interests and relevance. The article, ‘What to do when Stakeholders matter. Stakeholder Identification and Analysis Techniques’ \cite{Bryson2004}, provides an overview of the most widely used methods. Bryson offers various definitions of the stakeholder: in one of them, the stakeholder is defined as ‘any group or individual who can affect or is affected by the achievement of the organisation’s objectives’ \cite[p. 46]{Freeman1984}. When applying stakeholder analysis techniques to a research project, the term ‘research project’ replaces ‘organisation’. Two of the techniques presented could be a starting point for thinking systematically about stakeholders: the basic stakeholder analysis technique and the power versus interest grid.

\noindent\textbf{\emph{Basic stakeholder analysis technique.}}
The basic stakeholder analysis technique \cite[p.71-75]{Bryson1995} is a structured brainstorming method that may be used by individuals or groups. The process below is an adaptation of Bryson’s proposal \cite[p. 29-30]{Bryson2004}. 
\\

\noindent Brainstorm the list of potential stakeholders:
\begin{itemize}
    \item Prepare a separate flip-chart sheet for each stakeholder.
    \item Place a stakeholder’s name at the top of each sheet.
    \item Create a narrow column down the right side of each sheet and leave the column blank.
    \item For each stakeholder, in the area to the left of the narrow column, list the stakeholder’s expectations of the action research project.

\end{itemize}

\noindent Decide how satisfied the stakeholder is with the (planned) project. Use coloured dots to indicate a stakeholder judgement of ‘good’ (green), ‘fair’ (yellow) or ‘poor’ (red) in the empty right-hand column of their flip chart sheet.
\begin{itemize}
    \item Identify and record what may be done to quickly to address a stakeholder’s concerns. 
    \item Identify and record longer-term concerns with individual stakeholders and with the stakeholders as a group. 
\end{itemize}

\noindent Additional steps may be included, such as:
\begin{itemize}
    \item  Specify how each stakeholder influences the project.
    \item Decide what the project needs from each stakeholder.

\end{itemize}

\noindent\textbf{\emph{Power versus interest grid.}}
The power versus interest grid was first described by Eden and Ackermann \cite{EdenAckermann1988}.  Figure 3 presents the general idea:

\begin{center}  
\includegraphics[scale=.30]{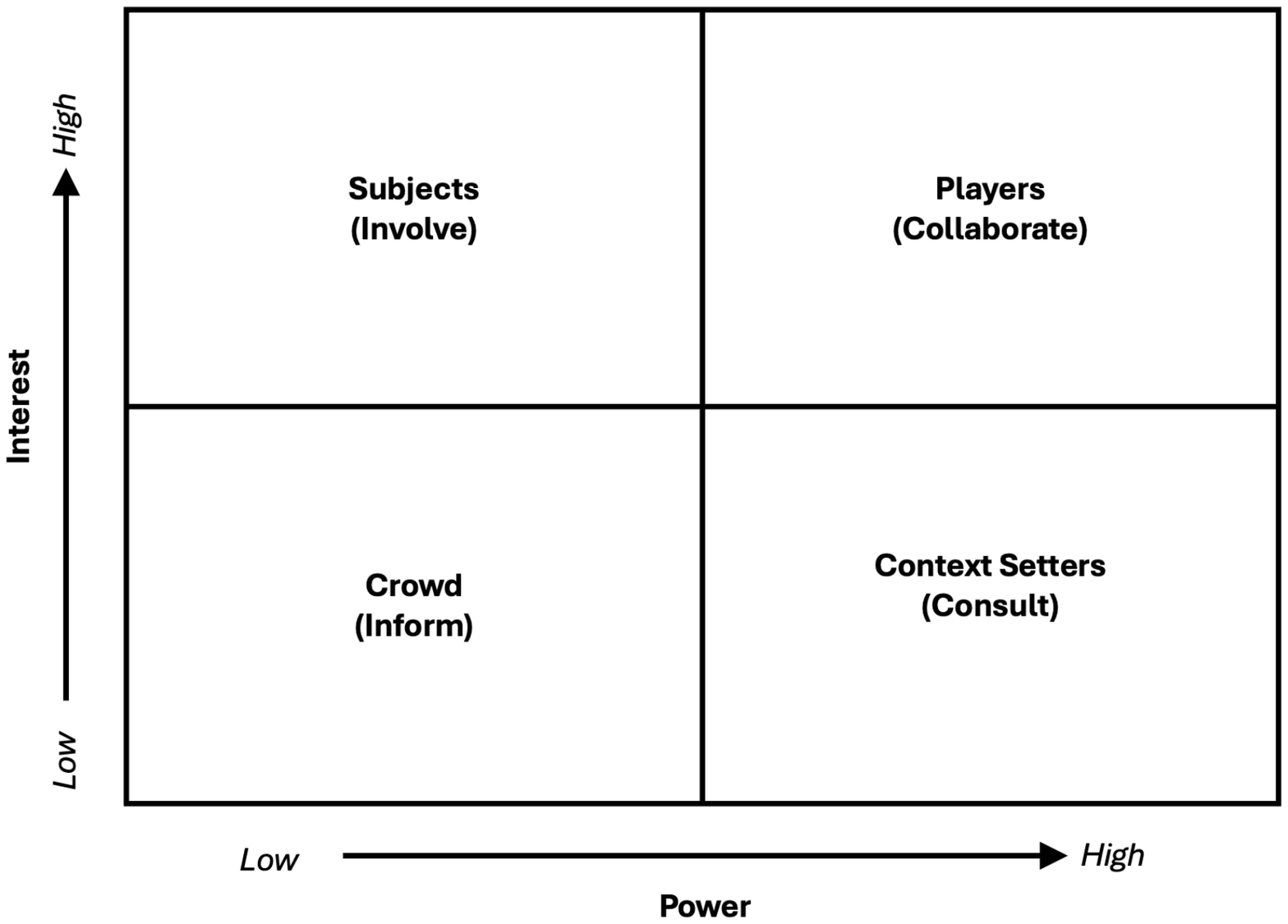}
\label{fig:3PowervsInterest}      

{\footnotesize \textbf{Fig. 3} Power versus Interest Grid \cite{EdenAckermann1988} }
\end{center}

The methods may be implemented by drawing the grid on a whiteboard and using sticky notes to position stakeholders, and move them according to the discussions of their interests in the project and their ability to influence the project, for example. You may use the stakeholders identified in the previous technique, or, if you start with the grid, you may brainstorm who the relevant stakeholders might be.

Both techniques give an idea of the organisational and inter-organisational setting for the action research project.
\\

\noindent\textbf{Informed Consent}

\noindent 
Informed Consent is a widely used concept in research with human subjects. It means that the people recruited for research need to agree to be subject to the research based on information that allows them to assess possible negative impact of the research on themselves. The informed consent can take many different forms. It can consist of presentations of the research that are signed by the project participants, it can also consist of the oral information about how data is handled in the beginning of the interview together with the assurance that the interviewee can ask the researchers to destroy the recording at any point in time. What form is adequate needs to be decided from case to case. 

In our practice, we have good experiences with written text containing: a short description of the project; a short description of the collaboration; a short presentation of the university and the research group maybe including names and background of the involved researchers; possible benefits for the participating developers; how the research might take place (observation, interviews, workshops); how the data is handled; and contact information of the researcher and, in case the researcher is a PhD student, the supervisor.
\end{tips}

\subsection{Responsibility and Interventions}
Action research does not aim to only understand software engineering practices, it also aims to intervene in and change software engineering practices. For many organisations today, software is either the main product or the core infrastructure of their business. In addition to ethical considerations that apply to other qualitative research, the deliberation surrounding, and implementation of change and interventions need to be carefully considered and executed responsibly.

As in ethnographic research, the impact of the disclosure of individual actions, the development team’s software practices, and how software is developed in a company may have repercussions for the individual, the team, or the company as a whole. As with qualitative research, the researchers need to first clear the citations and references to the work practices with the individual and team, before research findings are disclosed to management. Similarly, the company must be able to request that the research publications do not mention the company or that they hide specifics of the project (for a more comprehensive discussion of this dimension of research ethics, see the ‘Teaching and Learning Ethnography for Software Engineering Contexts’ chapter by Dittrich, Sharp and de Souza).

Action research is not only about understanding software development practices, but also about changing them. For example, if an intervention adversely affects an organisation’s ability to develop software, it may have a severe impact on economic outcomes. Therefore, the researcher’s interventions need to be carefully deliberated at both the team level and the company management level. The company and the team need to have the final say, as the researchers cannot assume the responsibility for financial losses. This also implies that informed consent forms for action research projects not only need to include information about handling data, but also must detail how the interventions are deliberated and implemented. Similarly, the steering committee needs to discuss both the informed consent and how interventions are chosen and implemented.

The research process may have effects on the organisation. Participatory action research emphasises the involvement of the people or team whose work practices are the subject of the intervention. This also needs to be anchored in the company management: changes that might affect the organisation are often ordained by management; when engaging in action research on software engineering practices, the company management must be informed of, and agree to, the relevant software teams also being involved in deciding about the intervention. This needs to be discussed and agreed on from the outset. The empowerment of the software team may in turn lead to internal tensions, for example, other teams may envy their opportunity to experiment with new methods. Similarly, the interventions may be seen as infringing on other organisational stakeholders’ interests.

Historical developments in a company need to be considered. A Danish project that was collaborating with various companies on CMM-based software process improvements found that in one company, previous software process improvement projects led to so much resentment among thecompany’s software developers that the research team had to switch to a problem-based approach to software process improvement \cite{iversen1999situated}. The next section focusses on the participatory deliberation of the intervention, as part of the action research.

\begin{question}{Exercise 2: Anchoring}
    
\noindent Extend the text developed for Exercise 1 with the following:
\begin{enumerate}
    \item a proposal for the team to collaborate with, and a proposal for the steering group
    \item a plan for the initial research aimed at understanding the current practices
    \item a stakeholder analysis that focusses on the possible effects of changes to the aspect of software development subject to the proposal
    \item an informed consent document that you may either share with the individual developers or present to the team, for example, as one of the slides that introduce you and your research

\end{enumerate}

\end{question}

\section{Deliberation – Anchoring Interventions and Improvements in the Organisation}
\label{sec:3Deliberation}
One of the core considerations of any action research project is the deliberation of the intervention. Especially when collaborating with industrial and public partners, the responsibility for the decision needs to remain with the organisation: changes to a company’s software development practices may have a substantial impact on the company. In the worst-case scenario, the quality of the software developed or the operations based on the software could jeopardise the company’s existence. Therefore, both the CMD approach \cite{CMD2008} and Staron’s chapter, which introduces action research, emphasise involving relevant parts of the organisation in the discussion and decision on the changes. This deliberation often takes place at two levels: when a collaboration between a university and an industry begins, the researchers involved, representatives from the company and practitioners agree on a general research focus. When the partners will be applying for external funding, this will be part of the j joint application.

One of the core considerations of any action research project is the deliberation of the intervention. Especially when collaborating with industrial and public partners, the responsibility for the decision needs to remain with the organisation: changes to a company’s software development practices may have a substantial impact on the company. In the worst-case scenario, the quality of the software developed or the operations based on the software could jeopardise the company’s existence. Therefore, both the CMD approach [17] and Staron’s chapter, which introduces action research, emphasise involving relevant parts of the organisation in the discussion and decision on the changes. This deliberation often takes place at two levels: when a collaboration between a university and an industry begins, the researchers involved, representatives from the company and practitioners agree on a general research focus. When the partners will be applying for external funding, this will be part of the j joint application.

Once the project has started, the general focus will be further developed: the initial empirical research may refine the understanding of the problem for both the researchers and the software developers involved. In parallel to this, a literature study by the researcher may yield additional ideas for how to address the organisation’s needs. These results should inform the discussion of, and decisions about the interventions.

The deliberation needs to bring together the results of the initial research, an agreement on the (refined) problem to be addressed, various possibilities for addressing the problem, and a decision regarding what and how to implement the change that is expected to address the problem. The presentation of initial research results should involve a discussion by the team members, regarding whether the team can confirm the results and, especially, the identified problems and challenges.

Sometimes, for example, when the initial research is comprehensive and has yielded new insights, or when various kinds of interventions need to be explored through further research, it may be necessary to organise a series of workshops, to give the researchers time to prepare their findings and input and the participants time to reflect on the discussion. If the intervention prioritised by the project is mission critical, further exploration. and perhaps even experimentation, may be needed, before a decision may be taken. The text box below presents an example of such a complex deliberation process.

\begin{svgraybox}
\noindent \textbf{Deliberation Example: Personas is not applicable}
\\
\label{example:Personas}
\noindent 
The article, ‘Personas is not applicable: Local Remedies Interpreted in a Wider Context’  \cite{Ronkkoetal2004} presents the lessons learned from a long and complex deliberation process. The initial research with the interaction design team of a mobile operating system and application developer identified a lack of ability to communicate the empirical grounding of interface design decisions. Personas were identified as a way to connect and discuss interaction design in relation to an abstract representation of the results of the empirical research, and also allow to bring in concrete user research data into the discussion.

The researcher did a literature survey and presented the academic perspective on Personas, which at the time of the research was a relatively new approach. The Interaction design team developed relevant personas to illustrate this approach. At the same time, Personas were explored through student projects at the university. The interaction design team decided on the Personas approach. However, to apply this approach, the team needed to convince the software development organisation and Marketing \& Sales. The software development team and the executive management were brought on board. However, Marketing \& Sales wanted to involve the customers. The customers consisted of competing mobile phone producers who teamed up to share the cost of operation system development. They could not agree on a common set of personas. In the end, this resulted in closing down the Personas project.

The analysis of the deliberation process resulted in the insight that a business setting might lead to that approaches that are fully valid from an interaction design and software engineering perspective are not applicable.

\end{svgraybox}

\noindent The example above also shows that the deliberation process has to take place at the team level, and must be anchored in management. Here, the steering committee discussed in the previous section plays an important role. The manager of the steering committee needs to be able to pinpoint which stakeholders to involve in the decisions in an intervention.

It is crucial for the researcher to understand that deliberation is not just about coming to a decision on the intervention, but that the discussions themselves are research data, and often provide an opportunity for a deeper understanding of the practices to be supported and aspects that may influence the applicability of certain methods. For example, in the hydraulic simulation case, the practitioners’ reluctance to develop comprehensive documentation of the software architecture resulted in a broader, interview-based study of software architecture awareness \cite{UnphonDittrich2010}. The interviews with tech leads and architects of a number of very diverse software products showed that software-product architects need to be informed of the developers’ changes, to guide the software developers and to be up to date on the challenges. In turn, these findings reveal an under-researched and neglected part of software architecture.

As action research takes place in cycles, the results of one intervention may lead to the identification of new issues and influence the decision on the next intervention. The following WMU example provides further insights into how multiple action research cycles developed during the research process. The example shows that the action research could not be planned ex ante, but that the researcher’s and the organisational stakeholders’ learning about needs and deliberating solutions in one action research cycle prompted new research cycles.

\begin{svgraybox}
    
\noindent\textbf{Case 3: World Maritime University (WMU) – Infrastructure and Methods for Participatory and End-User In-House Development}

\noindent\textbf{\emph{Background.}}
The action research was initiated to support shop-floor IT management practices at the World Maritime University (WMU). WMU operates with a capacity-building mandate under the United Nations special agency, the International Maritime Organisation (IMO). Shop-floor IT management practices are characterised by a close-knit collaboration between domain experts and IT professionals. Various shop-floor IT collaborative management groups were involved internally, to develop organisation-critical ICT systems, but were challenged by the need for an integrated infrastructure. The action research included three shop-floor IT development practices, including a Learning Management System and a Student Life-Cycle Management system to support the organisation’s capacity-building and educational programmes, and an organisation-wide contact database and an electronic forms system. Their composition, size and scope of development changed throughout the study, as needs and solutions evolved through action research. 

\noindent\textbf{\emph{Action research.}}
The Cooperative Method Development approach \cite{CMD2008} guided the action research, and was adapted to develop organisational IT management structures and processes, together with the organisational actors and IT professionals involved. The action research was a long-term project, and was carried out in three overlapping cycles, connected through their phases of understanding needs, and deliberating and evaluating improvement.

\noindent\textbf{\emph{Anchoring the deliberation in the organisation.}}
The various shop-floor IT collaborative management groups gradually became part of the research. The aim was to understand the needs and support improvements from the point of view of the shop-floor design constituencies that managed IT in the organisation. For example, in the first action research cycle, the researcher joined a team of faculty support staff, to develop a new web-based scheduler. This yielded situated insights into how a technical platform enables and constrains application development, and also how it is possible to work with participatory tools and techniques with the users. For example, a technical platform that supports only custom development had fewer provisions for rapid prototyping and end-user development. These realisations indicated that the technical platform could not be black-boxed for the users, but they needed to participate in its design. In the next action research cycle, the Student Life-Cycle Management system and contact database were included in the action research. Here, the focus was on how participatory tools and techniques could be used to involve users in the design of a new, organisation-wide Enterprise Resource Planning system.  

A range of various PD tools and techniques was applied, including functional analysis, to acquire an initial overview and gather requirements \cite{BødkerKensingSimonsen2004}; participatory observations (and stakeholders learning through overtaking work functions), together with story card summaries to acquire an in-depth understanding of current work practices \cite{BødkerKensingSimonsen2004, BeckAndres2004extreme, kyng1995} ; stakeholder workshops that used rich picture collages to identify key work practices and integrations, and also included mapping new solution scenarios; company visits, presentations by vendors, reviews and experimenting with prototypes \cite{BødkerKensingSimonsen2004}. The results showed how it was possible to involve the users in the design of the new ERP system, and how the PD tools and techniques supported the reorientation of shop-floor IT management practices, which were prompted by the technical and organisational integrations required by the new system. Thus, the results showed how it was possible to continue to empower users and apply shop-floor IT management practices when designing an integrated IT infrastructure. The final action research cycle targeted organisational IT management practices, and how users could jointly make decisions about their IT infrastructure and specific development projects through an IT steering committee. Improvements included both planning and decision-making that were connected to shop-floor IT management practices. The focussed action research cycles were not decided ex-ante, but gradually evolved according to the organisational needs.

\end{svgraybox}

\noindent The intertwined recognition of needs and deliberation, and the exploration of interventions were closest to each other in the Industriens Uddannelse case. The example below describes the close collaboration around the exploration of, and experimentation with, various representations and methods to support the use of familiar and new data sources.

\begin{svgraybox}
    
\noindent \textbf{Deliberation Example: IU –  Anchoring the Deliberation in the Organisation}

\noindent 
In the IU case, deliberation and intervention were often tightly intertwined; workshops that explored and designed tools and methods revealed the need for additional support, which was then addressed in a new workshop. An example is the third intervention of the IU project, which aimed to combine the learning from the previous interventions that addressed existing data practices and building design capabilities at the organisation, to further explore how the organisation could explore new data sources, and experiment with their usefulness. The intervention included a workshop series in which the researcher explored how members of the organisation could handle various data representations and their ability to co-design data. For example, one workshop focussed on how the members of IU could create ‘data searches’ as a way to explore how they look for data. Together, these workshops explored ways of ‘zooming out and zooming in’, an approach \cite{nicolini2012practice} to collectively understanding existing data practices. This combination of ‘macro- and micro-levels’ demonstrated that the education consultants worked primarily with existing data sources that were ‘ready at hand’, and made only limited use of data in new and innovative ways. 

This exploration motivated the implementation and co-design of IU’s Data Sphere, which was a tool that aimed to encourage all members of the organisation to consider and generate ideas for new data sources that the organisation could explore.
The intervention revealed an inherent focus on data, which supported the organisation’s consideration of data as something that may be used to innovate.

\begin{center}  
\includegraphics[scale=.65]{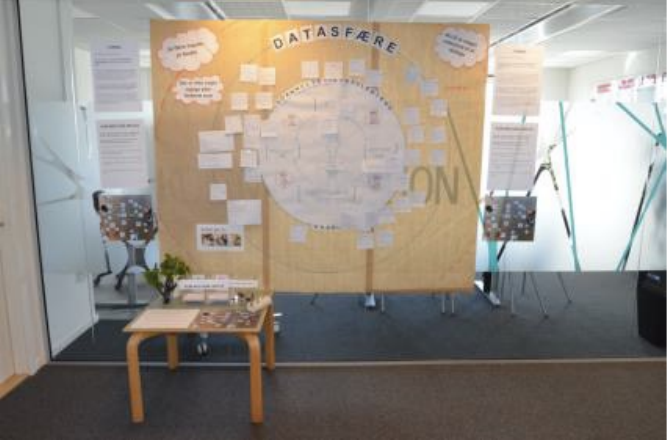}\\
\label{fig:4IU-Datasphere}      
{\footnotesize \textbf{Fig. 4} Case 2 – Datasphere }
\end{center}
\end{svgraybox}

\noindent The preceding examples show that, in many cases, deliberation consists of a series of activities. We recommend that beginning the process with a presentation of the results of the introductory research. This first step should give time for feedback from the team. A result might be a list of concerns or pain points that the team prioritises through some voting mechanism.
The next step could be the researcher presenting one or more approaches to addressing the pain points, which again may be subject to prioritisation.

As mentioned above, apart from the members of the development team, the management of the company’s development organisation also needs to be involved. Depending on the company culture, the agreement with the development team and its availability, management may take part in the deliberation workshops organised for and with the team. If that is not the case, or if ethical considerations are against it, the team’s decision needs to be confirmed by the steering committee. If necessary, the team and the steering group may involve other organisational stakeholders. Here, the initial mapping of relevant stakeholders and the initial research must inform the deliberation process.

The workshops in which the action research team discusses interventions, among themselves or with other parts of the organisation, need to be carefully prepared and implemented. The Toolkit Deliberation presents how to prepare, implement, and document workshops to support the practitioners involved and gather the data needed to support the research publications. Workshops may also be organised to fit the project and company culture. For example, when collaborating with an Agile team, the researcher could use one of the retrospectives as a deliberation workshop. Retrospectives are a construct for Agile development teams to reflect on their practices and to decide how to improve them. Similarly, project post mortems may be used to jointly reflect on implemented improvements, and evaluate them from a team perspective. 

\begin{tips}{Toolkit Deliberation}
\noindent\textbf{Workshop preparation, Implementation and Aftermath}

\noindent Especially in a software engineering context, the researcher collaborates with teams that consist of several people, and often a wider range of stakeholders is also involved, so workshops are often a good way to moderate a discussion among relevant stakeholders and prepare a decision.

Inspiration for workshops may be drawn from co-design and service design methods: In co-design and service design, the software, the collaboration among domain experts and the use of software applications is the subject of design. When researching action research, changes in work practices or the application of some methods and techniques are subject to the research.

Workshops need to be prepared, in order to foster useful discussions, and feel effective and efficient to the participants. As the emphasis is on the practitioners’ discussion and the input, presentations should be kept to a minimum.

Developing and leading a workshop is a skill that needs to be practised. We recommend the following process:

\noindent \textbf{\emph{Preparation.}}
You need to decide the purpose of the workshop and the intended outcome. If we use the improvement of a PhD study programme as an example, one goal of a first workshop could be identifying and agreeing on the core concerns and priorities. The next step would be to identify what needs to be in place for the participants to achieve this outcome. Next, activities should be selected that help to achieve these outcomes. This process may be recursive; especially when an activity itself requires other prerequisites. In the example of PhD programme-improvement, we could choose to brainstorm challenges of PhD studies and cluster them. Brainstorming could then be followed by a discussion of subgroups, which might lead to the elaboration of the root cause and the effects of the concerns. When the concerns are elaborated, this could be put to some form of vote.

The foregoing activities should be further refined and developed into a playbook, best presented as a table (horizontally formatted) with columns for time frame, activity, purpose, person responsible, approach and prerequisites. The time frame should be broken down to 5-minute intervals, where appropriate.

In parallel with the planning, time and place need to be chosen: A room needs to be reserved. Especially for longer workshops, coffee and catering need to be organised. Necessary materials, such as adhesive notes, paper and pens, and the like, need to be acquired.

\noindent \textbf{\emph{Implementation.}}
Usually, the workshop moderator would not be able to take field notes. If possible, workshops such as the one developed above should be audio- or video-recorded, as people may point, or otherwise communicate non-verbally.

Other group members or the supervisor might participate in the workshop, to support the researcher in case of various contingencies. For example, the workshop dynamics may develop in a direction that requires a deviation from the play book. Here, a second participant might help to manage the situation. If it is not possible to record the workshop, they could also take notes throughout the workshop.

Don't forget to prepare the room.

\noindent \textbf{\emph{Aftermath.}}
Before leaving the room, recordings need to be stopped and secured. The results of the workshop need to be documented: whiteboards and other physical results need to be photographed; if electronic media such electronic whiteboards were used, a screenshot needs to be taken. After the workshop, the room needs to be cleared according to the organisation’s guidelines.

Soon after the workshop, the participating researchers might meet for a debriefing, to discuss noteworthy events during the workshop and unexpected developments. The implications of the workshop’s outcomes for the next steps of the research need to be discussed.

\end{tips}

\begin{question}{Exercise 3: Planning and Implementing a workshop}
Plan a mid-term course evaluation using the Tool Kit Deliberation. Work in smaller groups and develop a playbook.

The playbooks can be presented in class and the class could vote on a specific playbook.

The team may then carry out their mid-term course evaluation using the chosen playbook.

\end{question}

\section{Visualisations - supporting mutual understanding and materialising complexity and change}
\label{sec:4Visualisations}

Action research often involves various forms of domain expertise. Apart from the researchers involved, an intervention might include other researchers, stakeholders with deep business knowledge, stakeholders with shop-floor expertise and software engineering stakeholders. These various domain experts bring valuable knowledge to the project, but might lack a shared language, which can lead to misunderstandings. Therefore, it is important to establish ways for the involved parties to develop a mutual understanding, to allow an intervention to develop and (ideally) generate the intended change. One way to go about this is by creating visualisations.

\begin{svgraybox}

\noindent \textbf{Visualisation Example:  WMU – Establishing mutual understanding through Rich Pictures}

\noindent 
Rich picture visualisations were an integral participatory technique when a design proposal for an Enterprise Resource Planning (ERP) system at WMU was being developed. An ERP system is an integrated infrastructure that connects an organisation’s administrative systems. Because of the complexity of connecting various technical systems and their associated work practices, ERP systems are difficult to design and implement \cite{Ciborra2000}. In the WMU case, the challenge was to support the users’ participation in the design process. The focus of the second action research cycle was to understand how rich picture workshops could be used in combination with other participatory design tools and techniques, for this purpose \cite{BødkerKensingSimonsen2004}. The action research cycle started with participatory observation. 

\begin{center}  
\includegraphics[scale=.65]{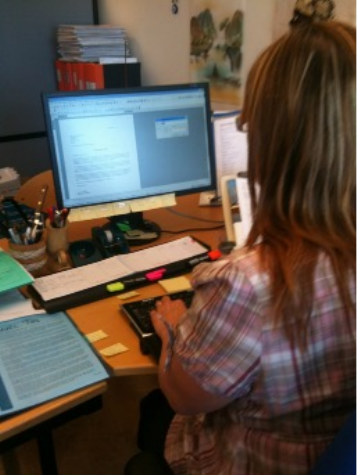}\\
\label{fig:5WMU-ObservingDomainExperts}      

{
\footnotesize \textbf{Fig. 5} Case 3 – Observing Domain experts }
\end{center}

\noindent The example of the image in Figure 5 shows how the researcher observed one of the administrative assistants, to gain an in-depth appreciation of her work and how she worked with the student life-cycle management system. The insights formed the basis for documenting the work practices and use of system support in what may be referred to as story-card summaries \cite{BeckAndres2004extreme}. The outcome of the story card summaries from various departments were then visualised as rich pictures, as seen in the image below.

\begin{center}  
\includegraphics[scale=.65]{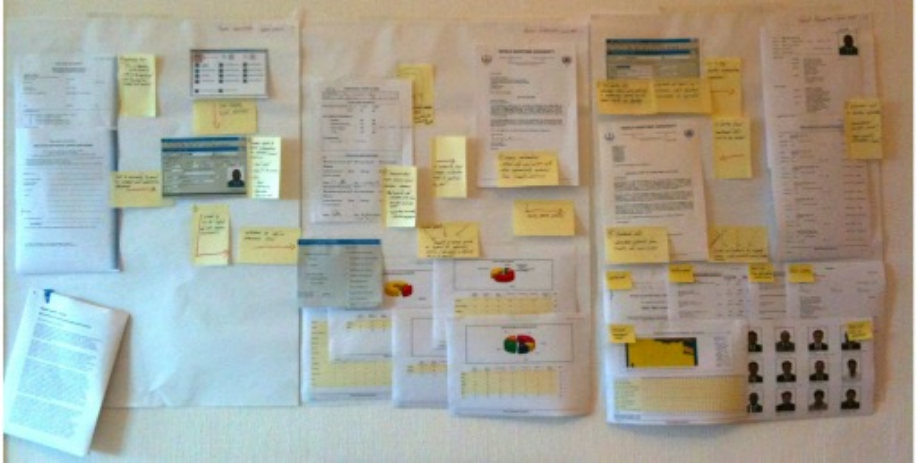} \\
\label{fig:6WMU-ExtendedRichPictures}      
{\footnotesize \textbf{Fig. 6} Case 3 – Extended Rich Pictures }
\end{center}

\noindent These rich pictures were first used so users from various departments could learn about each other’s dependence on system support, and how various systems were linked. Following this, the rich pictures were used to develop new ideas for improved integration, and workshops were arranged, so the users and the researcher could redraw possible connections between the systems. The outcome of the rich-picture-based design work was summarised in new story card summaries, and ERP system service providers were requested to use them as the basis for their presentation, so the users could follow these presentations, based on their previous collaborative work.

\end{svgraybox}

\noindent Visualisations are valuable artefacts in action research, from both the research perspective and the change perspective. From the research perspective, visualisations can support and deepen the researcher’s understanding of the industrial setting. Examples of such visualisations include representations of the organisational structure, the broader framework of the organisation or the IT infrastructure. The (often) concrete format of visualisations allows for instant ‘member checking’ as it is required for quality assurance in qualitative research \cite{RobsonMcCartan}: for example, the researcher may ask members questions about the organisation, about their understanding of a visualisation, and the ongoing adaptation and development of the visualisation. Moreover, a visualisation may act as a tool for reflection that supports the researcher’s and the organisation’s understanding of the often-complex reality which they must navigate, to achieve the intended change. From the change/learning perspective, visualisations may materialise the (sometimes) invisible action research intervention. Visualisations may help to demonstrate the impact of the intervention. Finally, if a visualisation is developed over time, it may also act as a way to document the development of the intervention.

\eject

\begin{svgraybox}
    
\noindent \textbf{Visualisation Example: SIM – Code comparison triggering integration of two software products}

\noindent As part of the fieldwork, the action researcher at SIM analysed the code of the software product subject to the collaboration. Part of that analysis was also a comparison of two sister products that simulated open and closed, one-dimensional hydraulic systems. As quite a few people became interested, the researcher displayed the visual representation of the results of this comparison in the company’s hallway. Shortly after, she observed developers of the two products standing in front of her printout and discussing the (minor) differences. The visualisations of the similarities and differences between the two products prompted a discussion that ultimately resulted in the integration of their computational core. Though the code analysis was a result of the researchers’ attempt to understand the architecture of the software, the visualisation also influenced the decisions to re-engineer both products.

\end{svgraybox}

\noindent 
As foregoing the examples show, the use of visualisations differs from traditional system specification. Their difference may be understood as the difference between the intention to define and to remind \cite{kyng1995}. Visualisation and participatory tools and techniques do not stand independent from the situation in which they have been produced, but are intended to remind (not define) the design team of work situations in particular need of computer support. They become boundary objects \cite{star1989institutional} that establish a field of interaction among various stakeholders, as they allow project members to relate heterogeneous perspectives. In all three examples presented in this section, the visualisations supported the communication among various members of the organisation and helped them communicate their respective perspectives, which were rooted in their specific work practices. They became collaborative artefacts. This is especially important in action research, where the researcher, together with users and other stakeholders, must understand needs and deliberate change. For this purpose, it is critical to share knowledge among individuals with different backgrounds, perspectives and motivations. Visualisations promote the verbalisation of implicit understandings as words and phrases, and the collaborative creation of new concepts through dialogue.

To promote dialogue among researchers, software developers and other stakeholders, the form of notation used must be well-chosen. Software engineers and researchers are familiar with formal state machines, and semi-formal notation, such as UML diagrams. When they use such notation, the focus is often on using the notation correctly, and the correct mapping of what should be depicted onto the model. Therefore, these forms of notation often limit the discussion to the correctness of the model. Though formal notation may play an important role, as in the comparison of the software products in the SIM case, we often use more flexible, expansive \cite{Engestrom1999, SeidelinDittrichGronval2020}  forms of visualisation, such as the rich pictures introduced in the toolkit part of this section. Such deliberately informal notations allow the addition of new perspectives to the picture, and provide space for opposing views of the same reality.

\begin{svgraybox}
    
\noindent \textbf{Visualisation Example:  IU – Mapping of the IT infrastructure}

\noindent As part of the fieldwork, the action researcher at IU created a visualisation – a map – of the organisation's IT infrastructure. The map was developed through meetings with IU’s only (at the time) external IT developer.

\begin{center}  
\includegraphics[scale=1]{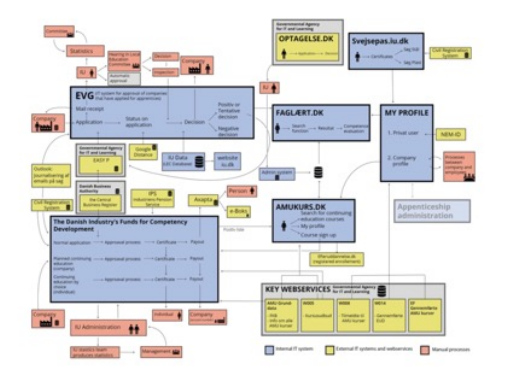} \\
\label{fig:7IU-MappingIUsITInfrastructure}      
{\footnotesize \textbf{Fig. 7} Case 2 – Mapping IU’s IT infrastructure }
\end{center}

\noindent The organisation had never developed a representation of their IT Infrastructure before, therefore this visualisation (see Figure 7) became a tool for explaining how the internal systems were integrated, and how and to what extent the organisation’s IT infrastructure – and by extension, data practices – depended on external stakeholders’ IT systems and web services. Over time, the visualisation was used as part of the reasoning for establishing an internal IT department and in this way improve the internal understanding of IT, and the extent to which the organisation’s IT infrastructure – and by extension, data practices – depended on external stakeholders’ IT systems and web services.

\end{svgraybox}

\noindent Finally, visualisations are a powerful tool for communicating research and research results. Figure 2 in the introduction of the IU example, which visualises the action research process, is the final version of a series of figures. An early version was developed at the beginning of the project; it was revised and used throughout the project to discuss progress and changes to the plans. The final version demonstrates the action research that was carried out. When presenting research results, visualisations may similarly play an important role: an example is Figure 10, below, which was first published in as part of a conference article \cite{SeidelinLeeDittrich2020}. The figure presents a diagram of the public sector arena for vocational education and training in Denmark. The diagram has been crucial to an understanding and the interdependence of data and cross-organisational collaboration, and to communicating the findings to the academic community.

Visualisations may be used to communicate research findings to industrial practitioners and domain experts, both within the project and in presentations to external industrial practitioners. The down side of visualisations is that good ones take time to create. Often, they are not created on the spot, but used as a tool to assist though, and accordingly are revised and developed over time. Also, not everybody is comfortable displaying his or her (lack of) drawing skills. However, we found in most cases, that the result is worth the effort. 

\begin{tips}{Toolkit Visualisation}

\noindent \textbf{Visualise – Check with relevant cooperators – Repeat} 

When complex circumstances are visualised, heterogeneous perspectives often need to be reconciled. In these cases, iterations of visualisation, checking the representations with relevant members of the organisation and repeating are important: that allows the stakeholders involved to express and explain their perspective. In two of the examples in this section – the WMU example of establishing mutual understanding through rich pictures and the IU example of mapping the organisation’s IT infrastructure – developing the visualisation was part of the research process. In the WMU example, the use of rich pictures was part of an intervention that demonstrated the scaling of participatory design methods at an organisational level. In the IU case, the visualisation was about mapping the data sources that were used to prepare and plan the intervention. For example, the map was used to identify the subject system of the first action research cycle.
\\

\noindent \textbf{Rich pictures and inspiration from design artefacts.}

\noindent It is challenging to create good and relevant visualisations, because they often need to convey complex information in a relatively simple manner, in order for it to resonate with the people and the project to which it relates. To begin with, we suggest looking at existing visualisation methods and, if necessary, adapting them to your needs. One of our favourites is the rich picture, a deliberately informal notation that is designed to be adapted and extended to the needs of the specific design or research purpose \cite{Mathiassenetal2012}. The set of elements below is only a starting point.

\begin{center}  
\includegraphics[scale=.65]{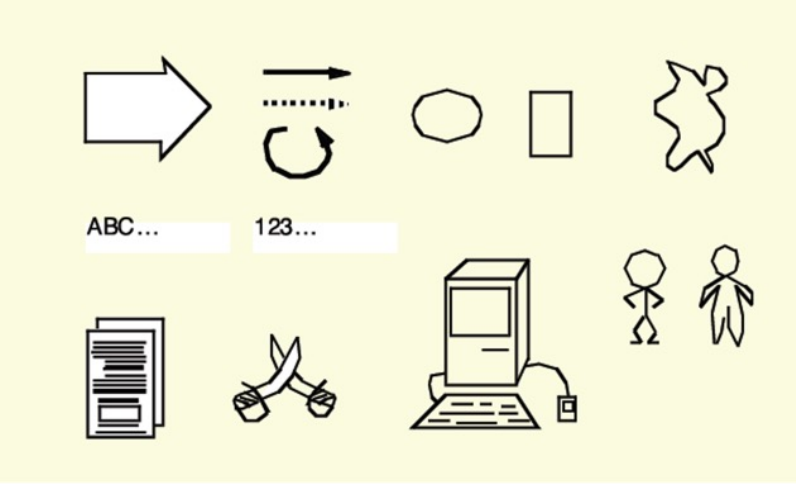} \\
\label{fig:8RichPictureBasicSymbolSet}      
{\footnotesize \textbf{Fig. 8} Rich Picture basic symbol set \cite{OOAD2000}}
\end{center}

\noindent If new visualisation elements are needed during a discussion with domain experts, they may be designed on the spot. Conflicting perspectives and unclear elements may be ‘put on the table’ or the whiteboard as crossed swords or clouds. Many user-centred design methods offer a number of graphic representations that typically may be flexibly adjusted, such as service blueprints, ecosystem mapping or timeline mapping.

\end{tips}

\begin{question}{Exercise 4: Rich Pictures}
    
\noindent Develop a rich picture of Stakeholders and Interests based on the Stakeholder analysis in Exercise 2. Explain the rich picture to a colleague and enhance it based on his or her questions about the case. Add the rich picture and a description of it to the stakeholder analysis and the project description.

\end{question}

\section{Challenges as learning opportunities }
\label{sec:5Challenges}
The core advantage, and also the main difficulty of participatory action research, in particular, is that it offers the opportunity for industrial practices and the setting in which software is developed to ‘talk back’ to the methods, techniques and tools proposed by the researcher: e.g. methods proposed by researchers may be controversially discussed by the software developers involved, or an intervention decided on does not happen for some reason. The challenges that emerge during the intervention often indicate blind spots in the established research. One example of this is the SIM case, below: Traditional software architecture research focusses on software architecture design and communicating the software architecture to the developers. When long-living software products are developed, communication about the difficulties faced by the existing architecture and the reasons why work-arounds occur are equally important. However, these needs tend to be neglected in the scientific discussion.

\begin{svgraybox}
    \textbf{Challenges Example: SIM – Being challenged to investigate the concept of ‘Walking Architecture’}

\noindent 
One of the main surprises for the collaborating software engineering researchers was the absence of documentation of the software architecture not only of the software product that was subject to the collaboration but to all software products maintained in the company. There technical leads would teach new developers as apprentices. They also doubted the usefulness of a document that described the structure, as it would soon be outdated.
The researchers wondered whether the distrust of software architecture documentation was due to the specificities of software product evolution, or whether it might be due to the developers being mainly applied mathematicians. To find out more, the researcher held a set of interviews with lead architects, tech leads and core developers of software products, from SMEs to product teams in multinational companies.
The interview study \cite{UnphonDittrich2010} confirmed that most of the product teams interviewed did not document their software architecture. The comparison with the only (open-source) software product that used a wiki to capture the software architecture indicated the rationale behind the lack of software architecture descriptions in other cases: The interviewee explained that every evening he needed to read the commit statements and review the changes to the wiki, so he could intervene, if a change was not in line with the rationale behind the design architecture. Without the extra effort, he would be unable to keep up to date. In other words, there might be good reasons for bad software architecture documentation that makes it necessary for the developers to consult the architect before changing the structure of the software.
In the final part of the field work, the team supported the action researcher in describing the core features of the architecture after it was re-engineered to support the simulation departments, if project-specific features needed to be included (Such customisations were needed rather frequently, and provided one line of input for the evolution of the software: If a project-specific feature was deemed relevant to other projects, it would often be revised and included in those products.). Here, there was no danger that the existence of documentation could result in an accidental evolution of the software architecture of the product.

\end{svgraybox}

In other cases, the challenges faced during deliberation and intervention indicate connections between subjects that are normally treated in different research communities. In section \ref{sec:3Deliberation}, we shared an example in which the deliberation resulted in a method welcomed by the interaction designers, software developers and technical management involved in the development that was not could not be applied in the organisation, because of the way the company collaborated with its customers. Here, the business model affected the applicability of an interaction design method. Such dependencies are neither subject to Interaction Design and Participatory Design, nor are they normally discussed in the economics literature. Similarly, in the WMU case: The example below describes how the action research evolved from focussing on co-design methods in organisational settings to the innovation of the technical base and the implementation of participatory IT Management structures. Again, the results indicate a connection between Co-Design and Information Systems that is normally not addressed in any of the related discourses.

\begin{svgraybox}
\textbf{Challenges Example: WMU – From Co-Design methods to redesign of the technical infrastructure and participatory IT Management}

\noindent IT development at the WMU was characterised by a strong emphasis on the expertise and support of the faculty and administration. The research project was launched to strengthen and further develop this Co-design tradition. The first action research cycle was implemented as part of the redevelopment of the teaching support, the scheduler of courses and rooms. One of the results showed that in many cases, implementing the interaction required by the domain experts would be impossible, given a university’s IT budget. At the same time, relevant parts of the functionality were implemented with standard software. As a result, the next action research cycle addressed the development of the technical basis of the IT infrastructure to allow for the use of standard office systems to edit data, such as Microsoft Excel™ for editing course and room schedules. Such a change in the technical infrastructure needed to be deliberated and determined by a computer committee that consisted of representatives from the IT department, various administrative departments and faculty. This led to exploring ways of making such complex changes that only very indirectly affect users comprehensible to members of the organisation who were not IT professionals. 
At the same time the WMU renewed its commitment to a co-determination approach to IT infrastructure development with the above-mentioned computer committee. During the process that led to the confirmation of a participatory IT Management strategy, it became apparent that, in order to effectively steer an increasingly complex IT infrastructure, better tools of IT Management were needed. These means were then subject to a third action research cycle. 
To answer this need, the topic of the PhD study developed from participatory design and end-user development to sustaining participatory design in an organisation 
\cite{BolmstenDittrich2015}.

\end{svgraybox}

\noindent 
Though the examples indicate action research’s potential for identifying and addressing blind spots in and between various research discourses, often these topics are difficult to handle. And, there is no one way to address such challenges. In the SIM case, the result was a side project that supported and triangulated the findings of the main study. In the SIM case, the insights shared by the practitioners and those based on the interview study informed the design of the intervention. In the other two cases, the difficulties encountered resulted in a change in the scope of the PhD dissertation. At the WMU, the second action research cycle focussed on the development of an infrastructure that allowed for the integration of standard systems for specific tasks. In the case in which Personas were not applicable, the subsequent research investigated how the communication between interaction designers and software developers improved \cite{ronkkoetal2008}. As the results questioned the role of methods, the reflection on methods and their roles was part of the final dissertation \cite{Ronkko2005}. 
In any case, both the PhD researcher and the supervisor needed to be flexible, to be able to respond to difficulties encountered, and to turn them into learning and research opportunities. However, this may be a challenge if a dissertation is involved. Students often have difficulties to estimate the complexity of a doctoral project. The supervisor has an important role in guiding the student, both academically and with respect to the practical aspects. For example, students are easily overwhelmed by the opportunity to make improvements, or underestimate the importance of navigating the organisational setting, which may present conflicting stakeholder interests. It may not be possible to determine the scope of the research before starting the empirical research; action research cycles may be an ongoing process, as the researcher and practitioner gain insights (through the interventions) into the work domain. Therefore, it is vital that the PhD student and the supervisor meet regularly, to handle the evolving action research in a manner that is compatible with the requirements for the degree. 

One of the tools that we used to manage the unfolding action research was debriefing, implemented in the research groups. Debriefings are an established part of qualitative research, and aim to formulate a first insight into the collected data, with the goal of preparing a systematic analysis and directing future data collection. Debriefings help to identify challenges early, and to explore them further, through discussions or through additional research. That way, a more nuanced understanding of the challenge and the root cause of the challenge may be obtained, and become the basis for research publications and further research.
\begin{tips}{Toolkit Challenges as Learning Opportunities}
\noindent \textbf{Debriefing}

\noindent 
Peer debriefing has long been discussed as a means of improving the quality of qualitative research \cite{CresswellMiller2000}. Debriefing is also used as an emergency response, for psychological stress management \cite{MagyarTheophilos2010} and for knowledge-sharing and learning related to future similar situations \cite{Conoscentietal2021}. In our practice, peer debriefing served both purposes: The debriefing took place primarily as part of supervisory meetings between the supervisor (as internal debriefer) and the researcher. However, in the IU case, the debriefing also took place between the researcher and a member of the organisation. These debriefings would often focus on recent research activities, and addressed organisational changes and/or development. The sessions were discussions between the two parties, who asked questions and reflected on new insights or things that could be changed or improved. 

In the cases of supervisor/researcher debriefings, the debriefer would ask questions to support the researcher’s reflection on these events. In many cases, the debriefing focussed on the research methods, and would result in decisions on next steps. However, special debriefing meetings focussed on the analysis and the substance of the research. They were scheduled either because of specific article writing projects, or because topics that came up in previous debriefing meetings. During these meetings, the debriefer typically asked about specific themes in the empirical work and explored the way the theme emerged in the field material, various aspects of the theme, related themes and so on. Such a debriefing meeting would typically end with a plan for further analysis or even with plans for additional data-collection. 

Debriefings help the researcher to cope with challenging situations, support knowledge-sharing, especially as a starting point or part of a collaborative analysis, and allow the identification of the scientific contributions of challenging developments and encounters.

\end{tips}

\begin{question}{Exercise 5: Debriefing}
\noindent Form groups of three in the classroom. Ask the students to spend 15 minutes writing down their reflections on their empirical research. The students could consider a challenging situation, a situation where they felt their role as an action researcher was difficult, a situation that felt out of the ordinary or the like. In the groups, one student will act as the debriefer, one as the researcher, and the third person will act as an observer. Each will take turns, to ensure that they experience debriefing from different perspectives. The person acting as the researcher should spend 5 minutes sharing the situation in question, and then the debriefer should ask questions. The debriefer may draw inspiration from ‘The 5 Whys technique’ \cite{Serrat2017}, which is an interactive interrogative method used to explore the root cause of a problem. The technique may support the students understanding of the reason for a ‘strange’ situation, or why their role was challenged at a certain point.

\end{question}

\section{Interventions as a stepping stone for creating sustainable change}
\label{sec:6Sustainability}

In addition to exploring the applicability of existing methods and developing new ones, action research may also aim to create sustainable change for the organisation with which the researchers are working, respectively the software development of that organisation. There are several reasons to aim for sustainable change. Continuous use is a more substantial proof for the usefulness of a method or tool than if the same only works under supervision of a researcher; another reason is to ensure that the organisation involved benefits from the resources invested in the research project.

So, what does sustainable change look like? In this section, we share empirical insights that illustrate how action research may generate sustainable change, and how to identify change in complex organisational settings. The aim is to help you through the process of identifying what constitutes change in your research project. 

Action research initiates organisational learning that may reach beyond the research project. Initiating change is at the core of the interventions that are part and parcel of action research. Furthermore, companies and organisations become involved in action research with the hope of improving their way of developing software. However, it may be very difficult to determine whether an intervention has effected sustainable change that persists beyond the end of the research collaboration. And, sometimes, the longer-lasting changes in an organisation are not directly related to the interventions of the action research. 

Action research does not have to explicitly aim for long-term change for the collaborating organisation. Especially for MSc or PhD students, evaluating a new method or tool with a single development project may be the only realistic goal. Action research is a time-consuming undertaking, and typically, students are constrained with regard to the time that is available for them to work on the dissertation, including carrying out empirical research, analysing outcomes and reporting the results. At the same time, even a project that focusses on a single application development may have a valuable outcome, both academically and for the organisation where the research has been carried out. Nguyen’s M.Sc. thesis at the World Maritime University is an example of a student carrying out one action research cycle, in this case, to design and test the usefulness of an educational technology solution for training seafarers during the COVID-19 pandemic\cite{Nguyen}. The prototype that the student developed yielded insights into how online, blended-learning modalities could be used, where the seafarers were unable to meet physically for their training. 

In many action research projects, the sustainable change in the organisation is both appreciated and supported. The interventions implemented together with the collaborating team focus in the first place on changing this team’s specific development practices. The interventions often address both the tools and the use of technology, and the team’s work practices. These local interventions may change the configurations of software development tools and techniques used by the organisation, and the organisational frameworks for, and methods of, software development. The research is geared to adding to the body of knowledge of software engineering, and eventually influencing the standard tools, methods and processes available to the organisation. We have formulated these levels of sustainable change as a framework in the toolkit box below.

\begin{tips} {Toolkit Sustainable Change}

\noindent\textbf{Understanding sustainable change}

\noindent 
The framework below structures the understanding of how action research learning and interventions for sustainable change take place in an organisational setting. The framework is inspired by an article on the impact of participatory development of educational technology on organisational processes \cite{BolmstenManuel2020} and builds on the work of Bødker et al. BødkerKensingSimonsen2004 and Pipek and Wulf \cite{PipekWulf2009}.

\begin{center}  
\includegraphics[scale=.38]{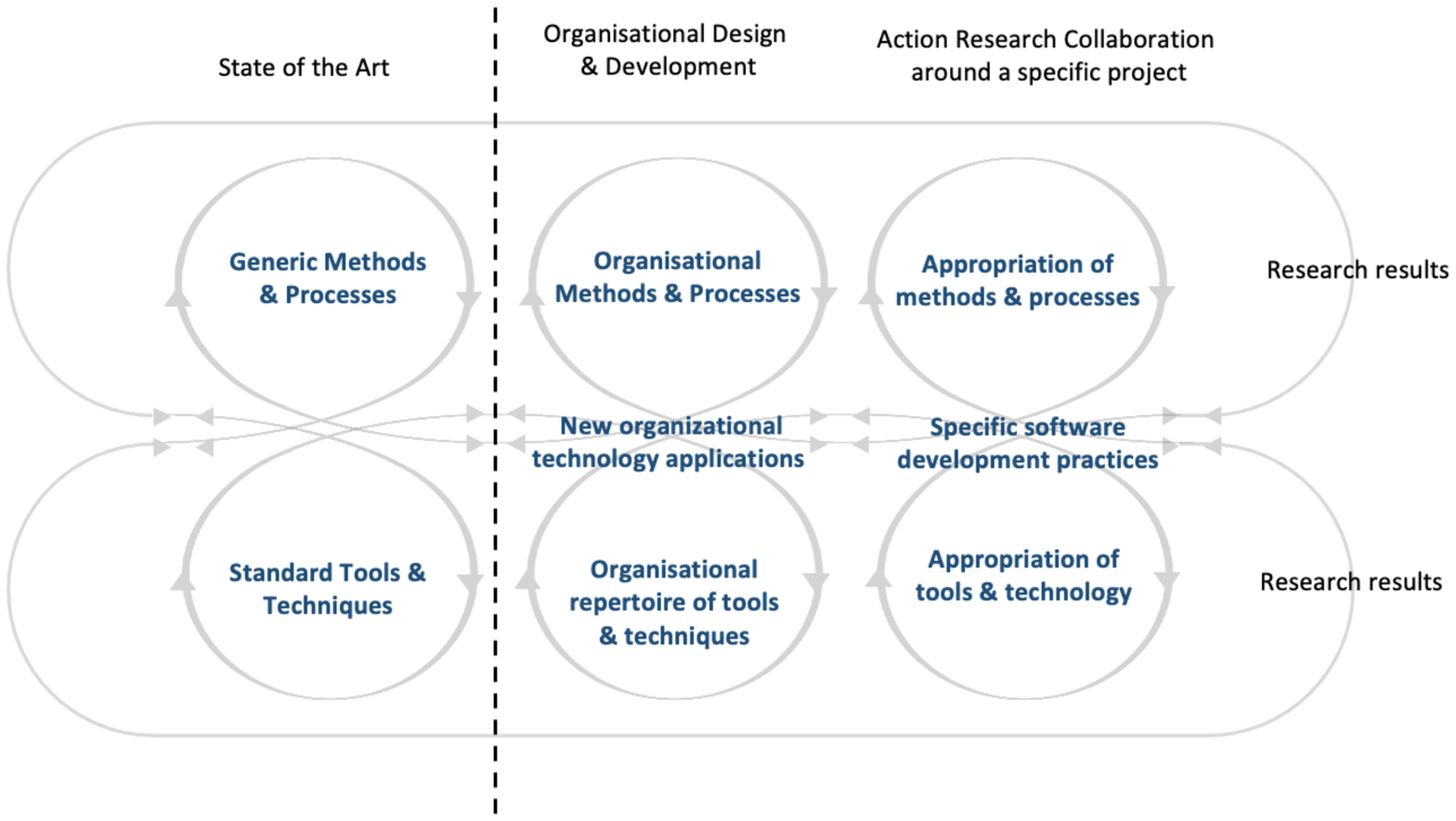}\\
\label{fig:9WMU-FrameworkForOrganisationalAnalysisOfActionResearchIntervention}      
{\footnotesize \textbf{Fig. 9} Framework for organisational analysis of Action Research Intervention}
\end{center}

\noindent 

The horizontal layers indicate the need for participatory knowledge development processes related to domain-specific knowledge (top horizontal layer) and technical knowledge (bottom horizontal layer), which, when combined, result in knowledge development related to the use of technologies (middle horizontal layer). This knowledge development, in turn, relates to the state of the art regarding techniques and methods (left-hand column), organisational repertoires of methods, processes, tools and techniques (middle column), and in-situ development and learning in a specific project (right-hand column).

Action research often starts at the individual project level. It may focus on either the appropriation and innovation of techniques and tools, the appropriation and innovation of methods and processes, and in many cases, it addresses how both the social and the technical aspects interact. As the deliberation example on page \pageref{example:Personas} indicate, the organisational history and structure influence the specific development practice and the outcome of an intervention. Action research may explicitly address the organisational layer, for example, by deploying innovative tools or introducing methods to a wider part of the. Even if the action research does not aim for innovation at the organisational level, it may influence the organisational level secondarily. The organisational level is informed by state of the art of software engineering methods and techniques communicated through professional networks or media, or through commercial and open-source tools. This state of the art also informs the action research through the researcher. Eventually, the research results will feed back into the body of knowledge.

The learning processes indicated in the framework by the looping arrows build on Susanne Bødker et al.’s concept of ‘knotworking’ \cite{BødkerDindlerIversen2017}. They show how ‘backstage’ development activities in the organisation are crucial to the sustainability of changes initiated by a specific action research project ‘onstage’. Given this understanding, action research focusses on the application and development of knowledge, both in the specific development setting and in connection to the learning processes across the various levels.

\end{tips}

\noindent 
All cases presented in this chapter serve as examples for creating sustainable change and innovation, both as intended parts of the research and in the form of secondary effects. In the SIM project, the focus was on introducing Agile software architecture practices that supported the continuous evolution of a software product. The introduction of software architecture concepts, especially the concept of product line architecture and the results of an exploratory comparison of the code bases of two sister products, had ripple effects on the organisation, beyond the specific context of the project. For example, it supported – if it did not prompt – the decision to integrate the two products. The project continued to develop lightweight methods and tools for maintaining the architecture of continuously-evolving software products, based on the study and intervention of the re-engineering project for the one-dimensional simulation engine. The example from the SIM project, below, highlights how changes to the development infrastructure may support the sustainability of organisational interventions.

\begin{svgraybox}

\noindent \textbf{Sustainable Change Example: SIM – Using artefacts to support organisational changes. (Build hierarchy)}

\noindent
The sustainability of the introduction of new methods and processes may be supported by materialising changes in the development infrastructure: In the SIM case, the researcher configured the build hierarchy so violations of the software architecture’s dependency constraints were discovered when the executable for testing was built \cite{Unphon2009a}. The team, especially the developers who joined the team, regarded this as support. They felt that the additional structural check served as a safety net that, together with the product’s new overarching architecture, allowed them to become productive much more quickly than before: historically, the company estimated that several months were needed for a new developer to understand the hydraulic simulation software well enough to contribute to its development; with the new architecture and the support through the built hierarchy, the time was reduced to a few weeks.

The configuration of the build system allowed the architecture to become a support for day-to-day development, and at the same time established affordances for discussing the architecture when needed. It also ensured that architecture evolution was implemented explicitly, as it required changes to the configuration of the development environment.

\end{svgraybox}

\noindent In the WMU case, the PhD researcher began by exploring participatory design methods for the development of one specific application. The experience of the concrete project led to a change in the technical framework used to develop teaching and learning support at WMU, which was trialled in a second development project. In parallel with this, it became apparent that the IT-management structure and tools needed to be reconsidered to handle such substantial changes to the IT infrastructure in an informed and accountable manner, which resulted in methodological and organisational interventions at that level. The new IT project-management approach was implemented and evaluated in a third development project. The example below illustrates how these dynamics unfolded.

\begin{svgraybox}
    \textbf{Sustainable Change Example : \textbf{WMU – Sustainable change as collaborative organisational learning} }

\noindent
The way in which the action research cycles in the WMU case alternated between the ICT system and method improvements at the local-level project with organisational-level improvements promoted sustainable change. The action research was anchored in the local level in the development of a Learning Management System, a Student Life-Cycle Management system and a Contact Database System that were used in everyday work by employees of the WMU. The action researcher collaborated closely with the associated domain experts. The software development that he contributed was validated in use. The local development iterations led to more substantial organisational level changes. One example of the connections established is illustrated by an early problem in the first action research cycle, when the researcher developed a new web-based scheduling system to replace an Excel-based scheduler. The administrator working with scheduling did not find it useful: ‘You were just not listening to me [...] that is not going to work’. The development needed to be redone. 

The problems encountered when developing the scheduler reoriented the action research and prompted two succeeding interventions at the organisational level: a.) The technical infrastructure that underpinned local software development was upgraded to support interfaces with office desktop systems. The upgrade also resulted in a revised architecture and a new programming language. b.) The upgrade, in turn, provided for new ways to develop software, as it enabled the rapid prototyping of new functionalities, which made it easier to test their usefulness with the domain experts. In this way, not only did the lessons learned from the scheduler application result in an improved version of the scheduler, but also in infrastructure and methods improvements.

Anchoring the interventions in the users’ and the organisation’s actual needs, and evaluating technical and methodological interventions in new software development projects, resulted in collaborative learning that, in turn, resulted in a long-term and sustainable change in the WMU case.

\end{svgraybox}

\noindent 

In the IU case, sustaining the change was part of the action research cycles from the very beginning. The first cycle focussed on exploring methods for designing with data related to a specific project. The second intervention and cycle targeted the organisation and their capability to work with design and innovation projects. The action researchers adopted methods from service design and supported various smaller-scale change projects in the organisation, so-called ‘service design micro-cases’ \cite{SeidelinSivertsenDittrich2020} . The final intervention and action research cycle focussed on supporting some members of IU to work more independently with the design of data \cite{SeidelinDittrichGronvall2020co}.

\begin{question}{Exercise 6: Reflecting sustainable change }

\noindent Using the toolkit framework of participatory and sustainable development processes, reflect on how your project, or an action research project you found in the literature, is positioned in relation to the framework. Think about the project’s focus and its action research cycles, and how those cycles relate to specific projects and the end-user, and organisational infrastructures and development methods. Furthermore, consider the project participants’ learning requirements, and the support that participatory tools and techniques may offer them.

\end{question}

\section{From Action to the Research Results}
\label{sec:7Writing}
Most action researchers might confirm that interaction with industrial practices is an adventure. However, the purpose of the research is the development of scientifically accountable results that may form the basis for future research, be it in the form of new action research or more controlled experiments. Publishing action research is challenging:

\begin{enumerate}
    \item As action research often relies heavily on qualitative empirical research, it presents the challenges that come with such flexible approaches: the research and analysis need to assure the trustworthiness of the research. Furthermore, as with all flexible research approaches, the research itself may result in changes to both the interventions and the empirical methods applied.
    \item Applying action research entails a conscious intervention in the observed practices. This entire intervention process needs to be regarded as data, which means all interactions with the team, including the analysis of the situation, deliberation on interventions, implementation of the interventions and evaluation are part of the research and need to be documented and becomes empirical data.
    \item To document the efficiency and effectiveness of an introduced method, additional research may be needed to establish a baseline before the intervention, to make it possible to compare this to the situation after the intervention. This baseline may be quantitative or qualitative. For example, if the agreed-on intervention is meant to improve the test efficiency of the development, the current test efficiency may be measured using quantitative measures. If the intervention should improve the communication between interaction designers and software engineers, the baseline may use qualitative interviews and the documentation of today’s development and communication practices as a baseline. The evaluation of the intervention needs to include corresponding methods. Often, the same quantitative measures may be applied to the new situation; using qualitative research, it may not make sense to ask the same questions as were asked previously, however, the same developers could probably be interviewed again when inquiring into the same aspects of the development.
    \item Software development is complex: smaller and larger groups of developers work in parallel; developers communicate with product managers, users and customers where appropriate; sometimes, the development is distributed over several sites or is even fully dispersed, with only virtual collaboration. The object of work, the software, is defined by program code, and based on and coordinated through a number of documents, which today reside in development environments. Aspects relevant to the research may not be readily apparent to the researcher. This means that the core action research may need to be complemented by suitable additional methods, such as source code and document analysis, or interviews with individual developers or entire groups.
\end{enumerate}

\noindent Newcomers and students who implement action research for the first time are often overwhelmed from the sheer amount of data. We recommend starting with the analysis and writing early in the process, to identify relevant research themes that underpin and are relevant to the research focus from which the action research started.

Analysing the initially-collected data and relating the findings to the relevant literature supports a reflection process that may also inform the deliberation and interventions. Also, later in the research, individual parts of the action research may be published as stepping stones. ‘Example From Action to Research Results: IU – Using publications as stepping stones during reflection’ demonstrates how writing articles may become part of the reflection process.

\begin{svgraybox}
    \textbf{From Action to Research Results Example: IU – Using publications as stepping stones in the reflective process}

\noindent 
The PhD dissertation based on the IU case resulted in 6 published papers (and two additional papers which were part of a related study). The papers were published during the course of the PhD programme, which was completed in 3.5 years. This required ongoing sequences of planning, conducting and analysing the research, and writing about the research findings. The papers built on each other. For example, the first paper focussed on understanding the current data handling in the organisation \cite{SeidelinDittrichGronvall2018data}. The analysis of the initial research presented two important research points: (a) that IU could be considered a ‘knowledge broker’, and that of cross-organisational collaboration needs to be considered in the action research, and (b) the concept of Human–Data Interaction. The publication was a stepping stone for reflection, as it helped us to understand that if we were to consider design with and of data in the IU setting, we could not consider the organisation in isolation. We also had to consider the organisation’s key stakeholders and collaborators. This informed the first intervention and influenced the design of the subsequent action research interventions. Over time, we developed a diagram that depicted the complex network in which IU navigates, and gives insight into the many stakeholders that need to be considered in relation to data handling, for example \cite{SeidelinLeeDittrich2020}.

\begin{center}  
\includegraphics[scale=.65]{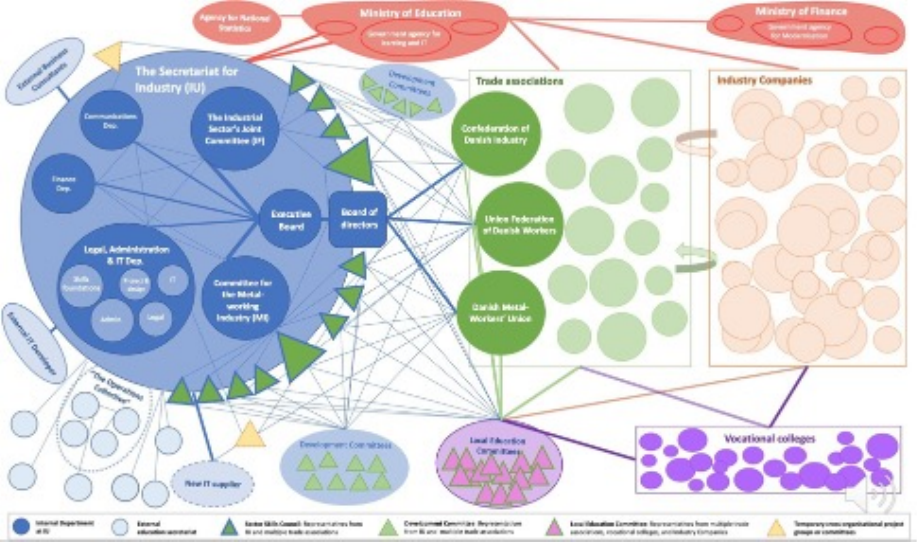} \\
\label{fig:10IU-DiagramOfThePublicSectorArenaForVocationalEducationAndTrainingInDenmark}      
{\footnotesize \textbf{Fig. 10} Case 2 – Diagram of the public sector arena for vocational education and training in Denmark \cite{SeidelinLeeDittrich2020}}
\end{center}
\end{svgraybox}

\noindent 
The themes of such early publications are often suggested by debriefing meetings with the research group or supervisor (See section 4 Challenges as learning opportunities above). Such debriefings also help to identify so-called ethnographic rich points (Agar, 1996), episodes encountered by the researchers that are experienced as unusual or surprising, and often lead to a deeper understanding of the research subject. In ethnographic and anthropological research, such rich encounters often provide a starting point for the analysis of the field material. These meetings then need to be followed by a systematic targeted analysis, for instance, the interaction analysis of the data design workshops in the IU example above. The toolkit introduces a thematic analysis process, and provides additional information about how to support data collection and analysis by looking for rich points.

The article, ‘Organisation matters: how the organisation of software development influences the development of product line architecture’ \cite{UphonDittrich2008} is an example of how debriefing can be used to support the identification of relevant research results: debriefing led to the understanding that not only do technical requirements and design impact the evolvability of the software under discussion, but the software-development organisation, practices, and business and use context matter as well. The results helped to shape both the intervention and the developing research \cite{Unphon2010,Unphon2009a}. 

When further exploring the results of a debriefing and/or ethnographic rich points, the research team may apply other suitable analytical methods. For example, when the team recognised that domain experts’ ability to design with data was apparent in specific interactions during the design workshops, interaction analysis \cite{JordanHenderson1995} was used to find, analyse and present examples of these exchanges \cite{SeidelinDittrichGronval2020}. Experience with a variety of social science methods is an advantage. In other cases, as in the WMU example below, a theme emerged that had not yet appeared in the core of the fieldwork, but had appeared as a border case at the fringes of the main research. In such situations, targeted empirical research, such as interviews or a focussed retrospective, may be designed, to triangulate with the existing field work. A second example is the above-mentioned article inspired by the SIM case, ‘Software architecture awareness in long-term software product evolution’ \cite{UnphonDittrich2010}.

\begin{svgraybox}
    \textbf{From Action to Research Results Example: WMU – Implementing additional research to triangulate field work}

\noindent The WMU case shows how triangulating the field work became important as the researcher was closely involved in the organisation. On the one hand, a close relationship between the researcher and the organisation provides an opportunity to acquire an insider’s understanding of the development dynamics and to contribute to making useful improvements as part of the action research; on the other hand, a common concern that was accentuated in the WMU case is that the researcher is ‘going native’, and does not have enough reflective space or distance from the organisation to analyse the events taking place. Therefore, it became important that the supervisor supported the researcher during the field work. The contact database and electronic forms were two of the systems that were part of several action research cycles at WMU. An administrative assistant managed this development. As an end-user developer, she carried out the technical development herself. The role of the researcher was to support her by making improvements to the underlying technical infrastructure. To understand her work and deliberate improvements, the researcher used participatory observation and participatory design techniques. 

When collecting data about End-User Development practices this became a problem: important aspects of her practices did not become explicit, as the administrator expected the researcher to know these aspects. To address these challenges, the supervisor interviewed the administrative assistant separately. In this way, several means of documenting the field work were employed, which triangulated each other and could be used for the subsequent analysis and report on the research.
\end{svgraybox}

\noindent 
Very few reports of full action research cycles have been published in core software engineering journals and conferences. One example from our research may be found in the article, ‘Organisational IT managed from the shop floor: Developing participatory design on the organisational arena’ \cite{BolmstenDittrich2015}. Reporting action research in articles is often challenging, as the complexity of the research may not be easily condensed into 10,000 or even 20,000 words. Therefore, especially when PhD students are part of the research, the full account is often presented as a dissertation \cite{Seidelin2020, Bolmsten2016, Unphon2010}. In their article, ‘Style Composition in Action Research Publication’, Mathiassen et al. provide a systematic literature analysis of action research publications in the Information Systems field, with respect to how the arguments and the contributions are presented \cite{Mathiassenetal2012}. They use the scheme adapted in Table \ref{tab:1ReportingStyle} to categorise the articles they found.

\begin{table}[!t]
\caption{Style Composition in Action Research Publications, adapted from Mathiassen et al. \cite[p. 351]{Mathiassenetal2012}
}
\label{tab:1ReportingStyle}       
%
%
\begin{tabular}{p{3cm}p{3cm}p{5,3cm}}
\hline\noalign{\smallskip}
Premisses Style & Inference Style & Contribution Style \\
\noalign{\smallskip}\svhline\noalign{\smallskip}
\textbf{Practical:} Argument based primarily on challenges in software engineering practices

\textbf{Theoretical:} Argument primarily based on challenges in software engineering research and theory
& 
\textbf{Inductive:} Argument primarily grounded in evidence from the software development practices that are subsequently related to concepts from the research context

\textbf{Deductive:} Argument primarily grounded in concepts from the research context, which are subsequently validated or illustrated by evidence from the software engineering practices
&
\textbf{Experience report: }Argument for insights from software development practices that may lead to research contributions

\textbf{Field study:} Argument for a primary contribution to improve the knowledge about certain kinds of development practices

\textbf{Theoretical development: }Argument for a primary contribution to a more general theory about software development

\textbf{Problem-solving method: }Argument for a primary contribution concerning how to address certain kind of software development problems

\textbf{Research method: }Argument for primary contribution to extend the knowledge of research methods and their application
\\

\noalign{\smallskip}\hline\noalign{\smallskip}
\end{tabular}

\end{table}

In many action research projects, researchers and practitioners collaborate closely, sometimes so closely that the practitioners are part of the research development. If the developers involved are willing and able to put in the additional effort, they may even co-author some of the articles. As in most cases the practitioners involved do not have research skills, the team’s researchers will still take the lead. To warrant co-authorship according to academic standards, all co-authors must have been involved in the reported research, and should have participated in drafting and writing the article. The former would often present no problem in action research. To invite industrial software developers into the writing process, the researchers need to at least discuss outline and argumentation with the practitioners, provide the opportunity for the practitioners to contribute to the writing, and allow time for the practitioners to review and add to at least one complete version of the draft. This may require more discipline with respect to time commitments during the writing process than usual in many academic contexts. Examples of action research articles include \cite{Ronkkoetal2004, ronkkoetal2008}.

\begin{tips}{Toolkit From Action to the Research Results}
\textbf{Thematic analysis}

\noindent Thematic analysis is a common analytical method in the social sciences, and may be used during both the problem analysis and the evaluation of the results in action research. Typically, a thematic analysis starts with identifying topic or coding recorded empirical material (e.g. from interviews, focus groups or workshops), then proceeds with organising the empirical material into themes. This provides a basis for presenting the findings as a quotation-rich narrative. The coding may use a predefined scheme based on the research objectives and questions, or be based on specific in-situ insights (or a combination of both). For a more in depth discussion see the chapter on qualitative analysis by Treude in this volume. The example in the figure from the WMU case shows how codes with annotations are clustered into themes. At a basic level, simple tools such as sticky notes may be used to code the material, but qualitative data analysis software is useful for systemising the analysis, especially if the material is comprehensive.

\begin{center}  
\includegraphics[scale=.65]{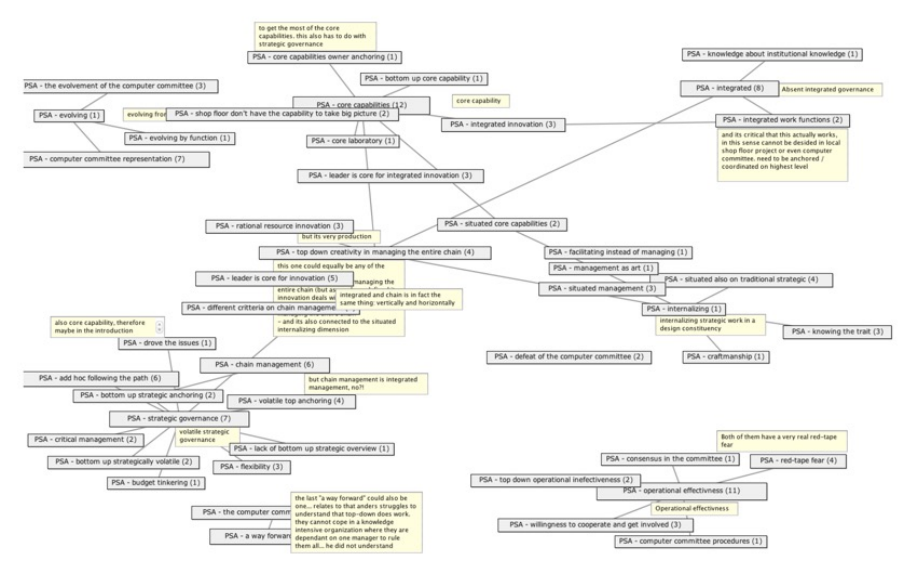} \\
\label{fig:11WMU-ExampleOfCodesStructuredIntoThemesFromTheWMUCase}      
{\footnotesize \textbf{Fig. 11} Case 3 – Example of codes structured into themes from the WMU case. }
\end{center}

\noindent\textbf{Ethnographic rich points}

\noindent One approach to practically guiding the data collection and analysis is to look for rich points. Rich points are those surprises and insights that the researchers (and the participants) encounter during the action research \cite{Agar1996}. They fuel ethnographic research, which is recommended for both the first needs analysis phase and final evaluation phase of the CMD approach. These may inform the coding in two ways. 

\begin{itemize}
    \item Encyclopaedic shared knowledge ethnography: What may be thought of as classic, encyclopaedic, ‘shared knowledge’ ethnography is referred to by Agar  as ‘disk contribution’ \cite[p. 12-16]{Agar1996}. The goal is to find common threads, in the sense of ‘patterns’ or ‘value configurations’ that appear in several cases where the local and individual complications move to the background.
    \item Narrative ethnography: Narrative ethnography may be referred to as a ‘pick contribution’ [ibid]. It concerns the practices of everyday life, the way those practices are built on shared knowledge, and all the other things that are relevant to the moment. It foregrounds rich points and their connection across domains and levels, rather than aiming to translate them into a generalisable pattern \cite[p. 9]{Agar2008}. This provides the capacity to identify individual ‘complications and contradictions, not as evidence for the encyclopaedia, but as problems to explain in their own right’ \cite{Agar1996}.
\end{itemize}

According to Agar’s propositions, the foregoing two options are not mutually exclusive; instead, they can be and are often used together. A narrative account that keeps close to the details of an everyday development is combined with seeking common patterns in the empirical material. The latter may serve as the foreground or background to the former, but the idea is not to keep individual complications in the foreground while seeking common patterns among cases.

\end{tips}
\begin{question}{Exercise 7: Prepare for publication}

\noindent Discuss the Premise Style, the Inference Style and the Contribution Style (Table \ref{tab:1ReportingStyle}, see also Mathiassen et al.’s article \cite{Mathiassenetal2012}) for the action research project you have been developing in some of the  exercises, and append the discussion and result to the project description.

\end{question}
\section{Conclusion}
\label{sec:8Conclusion}
Action research projects rarely play out exactly as originally envisioned. This chapter demonstrates how the challenging uncertainty of action research may be mitigated to allow the valuable research process and insights to be completed. Action research allows us to explore the appropriation and innovation of tools, methods and processes in real-world software development settings. It broadens our understanding of the factors that influence the adaptability of methods and helps identify how to improve the methods, to address new challenges – such as the co-design of data in the IU case, and the interaction between software developers, domain experts and end-user developers at the World Maritime University – and the adaptation of existing methods to fit specific kinds of settings: the SIM case yielded a better understanding of how software architecture methods need to be fitted into software product development and evolution. Action research helped us to extend the research beyond an understanding of software development environments through ethnography or grounded theory, for example, and it allowed to develop methods in situ, instead of ideating them and trying them out with the help of  student projects.

Though we continue to apply action research, we cannot claim that it is unproblematic. In this chapter, we aimed to address some of its difficult aspects: the interactions with an organisation that is developing itself independently of the research; the participating organisation’s reactions to the proposed intervention, which might result in changes to the research plans, and might even lead to a change in the research question; and last but not least the publication of action research results. We also share tools, techniques and processes that we developed or adopted from others’ work, and that helped us to meet the challenges of action research. We hope that after reading this chapter, the readers will be inspired and understand some of the concerns of being involved in action research and how to address them.

\begin{acknowledgement}
Writing this chapter provided us with the opportunity to reflect on the didactic aspect of our teaching and learning about action research. In particular, it enabled us to collect and reflect on the tools we use to address the sometimes challenging interactions with industrial software and IT development. We would like to thank the editors for providing us with this challenge. The reviewers of this chapter helped us to develop our conviction that this unusual piece would eventually fall into place. 
Last but not least, we would like to thank the collaborators in the various projects who contributed to our learning and supported the further development of our understanding and practice of action research.

\end{acknowledgement}

\bibliographystyle{spmpsci}
\bibliography{ARinIndustry}

@book{Agar1996,
   author = {Agar, Michael H.},
   title = {The professional stranger: An informal introduction to ethnography},
   publisher = {Academic Press},
   address = {San Diego},
   DOI = {papers3://publication/uuid/96F108B0-0BDC-4D1B-AAF2-BB963B9F08E9},
   url = {http://www.citeulike.org/group/268/article/221525},
   year = {1996},
   type = {Book},
}

@article{Agar2008,
   author = {Agar, Michael},
   title = {Culture: Can You Take It Anywhere?},
   journal = {International Journal of Qualitative Methods},
   volume = {5},
   number = {2},
   pages = {1},
   DOI = {papers3://publication/uuid/07429980-CBE1-45C1-8E90-B6E59B9529CA},
   url = {http://www.ethknoworks.com/},
   year = {2008},
   type = {Journal Article},
}

@book{BeckAndres2004extreme,
  title={Extreme programming explained: embrace change},
  author={Beck, Kentand Andres, Cynthia},
  year={2000},
  publisher={addison-wesley professional},
}

@book{BødkerKensingSimonsen2004,
   author = {Bødker, Keld and Kensing, Finn and Simonsen, Jesper},
   title = {Participatory IT design: designing for business and workplace realities},
   publisher = {MIT Press},
   address = {London},
   year = {2004},
   type = {Book},
}

@article{BødkerDindlerIversen2017,
   author = {Bødker, Susanne and Dindler, Christian and Iversen, Ole Sejer},
   title = {Tying Knots: Participatory Infrastructuring at Work},
   journal = {Computer Supported Cooperative Work (CSCW)},
   volume = {26},
   number = {1},
   pages = {245-273},
   ISSN = {1573-7551},
   url = {https://doi.org/10.1007/s10606-017-9268-y},
   year = {2017},
   type = {Journal Article},
}

@phdthesis{Bolmsten2016,
   author = {Bolmsten, Johan},
   title = {Sustaining Participatory Design in the Organization: Infrastructuring with Participatory Design},
   school = {IT University of Copenhagen},
university = {IT University of Copenhagen},
   type = {PhD Thesis},
   year = {2016},
}

@inproceedings{BolmstenDittrich2011,
   author = {Bolmsten, Johan and Dittrich, Yvonne},
   title = {Infrastructuring When You Don't - End-User Development and Organizational Infrastructure},
   booktitle = {End-User Development: Third International Symposium, IS-EUD},
   editor = {Costabile, Maria Francesca and Dittrich, Yvonne and Fischer, Gerhard and Piccinno, Antonio},
   publisher = {Springer Berlin Heidelberg},
   volume = {6654},
   pages = {139-154},
   ISBN = {978-3-642-21529-2},
   url = {http://dblp.uni-trier.de/db/conf/iseud/iseud2011.html\#BolmstenD11},
    year={2011},
   type = {Conference Proceedings},
}

@inbook{BolmstenDittrich2015,
   author = {Bolmsten, Johan and Dittrich, Yvonne},
   title = {Organisational IT managed from the shop floor: Developing Participatory Design on the organisational arena},
   booktitle = {Designing Socially Embedded Technologies in the Real-World},
   publisher = {Springer, London},
   pages = {383-417},
   year = {2015},
   type = {Book Section},
}

@article{BolmstenManuel2020,
   author = {Bolmsten, Johan and Manuel, Michael Ekow},
   title = {Sustainable participatory processes of education technology development},
   journal = {Educational Technology Research and Development},
   volume = {68},
   number = {5},
   pages = {2705-2728},
   ISSN = {1556-6501},
   DOI = {10.1007/s11423-020-09803-3},
   url = {https://doi.org/10.1007/s11423-020-09803-3},
   year = {2020},
   type = {Journal Article},
}

@book{Bryson1995,
   author = {Bryson, J.M.},
   title = {Strategic Planning for Public and Nonprofit Organizations: A Guide to Strengthening and Sustaining Organizational Achievement},
   publisher = {Jossey-Bass Publishers},
   ISBN = {9780787901417},
   url = {https://books.google.se/books?id=yydHAAAAMAAJ},
   year = {1995},
   type = {Book},
}

@article{Bryson2004,
   author = {Bryson, John M},
   title = {What to do when stakeholders matter: stakeholder identification and analysis techniques},
   journal = {Public management review},
   volume = {6},
   number = {1},
   pages = {21-53},
   ISSN = {1471-9037},
   year = {2004},
   type = {Journal Article},
}

@book{Ciborra2000,
   author = {Ciborra, C.},
   title = {From control to drift: the dynamics of corporate information infastructures},
   publisher = {Oxford University Press},
   address = {USA},
   year = {2000},
   type = {Book},
}

@article{Conoscentietal2021,
   author = {Conoscenti, Elena and Martucci, Gennaro and Piazza, Marcello and Tuzzolino, Fabio and Ragonese, Barbara and Burgio, Gaetano and Arena, Giuseppe and Blot, Stijn and Luca, Angelo and Arcadipane, Antonio},
   title = {Post-crisis debriefing: A tool for improving quality in the medical emergency team system},
   journal = {Intensive and Critical Care Nursing},
   volume = {63},
   pages = {102977},
   ISSN = {0964-3397},
   year = {2021},
   type = {Journal Article},
}

@article{CresswellMiller2000,
   author = {Creswell, John W and Miller, Dana L},
   title = {Determining validity in qualitative inquiry},
   journal = {Theory into practice},
   volume = {39},
   number = {3},
   pages = {124-130},
   ISSN = {0040-5841},
   year = {2000},
   type = {Journal Article},
}

@inbook{Dittrich2002,
   author = {Dittrich, Yvonne},
   title = {Doing Empirical Research on Software Development: Finding a Path between Understanding, Intervention, and Method Development},
   booktitle = {Social Thinking—Software Practice},
   editor = {Dittrich, Yvonne and Floyd, Christiane and Klischewski, Ralf},
   publisher = {The MIT Press},
   pages = {243-262},
   ISBN = {9780262271783},
   DOI = {10.7551/mitpress/6308.003.0016},
   url = {https://doi.org/10.7551/mitpress/6308.003.0016},
   year = {2002},
   type = {Book Section},
}

@article{CMD2008,
   author = {Dittrich, Yvonne and Rönkkö, Kari and Eriksson, J. and Hansson, Christina and Lindeberg, Olle},
   title = {Cooperative method development: Combining qualitative empirical research with method, technique and process improvement},
   journal = {Empirical Software Engineering},
   volume = {13},
   number = {3},
   pages = {231-260},
   year = {2008},
   type = {Journal Article},
}

@book{EdenAckermann1988,
   author = {Eden, C. and Ackermann, F.},
   title = {Making Strategy: The Journey of Strategic Management},
   publisher = {SAGE Publications},
   address = {London},
   ISBN = {9780761952244},
   url = {https://books.google.se/books?id=rveTQgAACAAJ},
   year = {1998},
   type = {Book},
}

@article{Engestrom1999,
   author = {Engeström, Yrjö},
   title = {Expansive Visibilization of Work: An Activity-Theoretical Perspective},
   journal = {Computer Supported Cooperative Work (CSCW)},
   volume = {8},
   number = {1},
   pages = {63-93},
   ISSN = {1573-7551},
   year = {1999},
   type = {Journal Article},
}

@phdthesis{Eriksson2008,
   author = {Eriksson, Jeanette},
   title = {Supporting the Cooperative Design Process of End-User Tailoring},
   university = {Blekinge Institute of Technology},
    school = {Blekinge Institute of Technology},
   type = {Doctoral thesis},
   year = {2008},
}

@book{Freeman1984,
   author = {Freeman, R.E.},
   title = {Strategic Management: A Stakeholder Approach},
   publisher = {Pitman},
   address = {Boston, MA},
   ISBN = {9780521151740},
   year = {1984},
   type = {Book},
}

@article{Ridleyetal2023,
   author = {Jones, Ridley and Seidelin, Cathrine F and Neang, Andrew B and Lee, Charlotte P},
   title = {Lessons Learned from a Comparative Study of Long-Term Action Research with Community Design of Infrastructural Systems},
   journal = {Proceedings of the ACM on Human-Computer Interaction},
   volume = {7},
   number = {CSCW1},
   pages = {1-35},
   ISSN = {2573-0142},
   year = {2023},
   type = {Journal Article},
}

@article{Hayes2018design,
  title={Design, action, and practice: Three branches of the same tree},
  author={Hayes, Gillian R},
  journal={Socio-informatics},
  pages={303--318},
  year={2018},
}

@article{iversen1999situated,
  title={Situated assessment of problems in software development},
  author={Iversen, Jakob and Nielsen, Peter Axel and Norbjerg, Jacob},
  journal={ACM SIGMIS Database: the Database for Advances in Information Systems},
  volume={30},
  number={2},
  pages={66--81},
  year={1999},
  publisher={ACM New York, NY, USA},
}

@article{JordanHenderson1995,
   author = {Jordan, Brigitte and Henderson, Austin},
   title = {Interaction Analysis: Foundations and Practice},
   journal = {Journal of the Learning Sciences},
   volume = {4},
   number = {1},
   pages = {39-103},
   ISSN = {1050-8406},
   year = {1995},
   type = {Journal Article},
}

@article{kyng1995,
   author = {Kyng, Morten},
   title = {Making representations work},
   journal = {Communications of the ACM},
   volume = {38},
   number = {9},
   pages = {46-55},
   year = {1995},
   type = {Journal Article},
}

@article{MagyarTheophilos2010,
   author = {Magyar, Joanne and Theophilos, Theane},
   title = {Debriefing critical incidents in the emergency department},
   journal = {Emergency Medicine Australasia},
   volume = {22},
   number = {6},
   pages = {499-506},
   ISSN = {1742-6731},
   year = {2010},
   type = {Journal Article},
}

@article{checkland2007action,
  title={Action research: its nature and validity},
  author={Checkland, Peter and Holwell, Sue},
  journal={Information systems action research: An applied view of emerging concepts and methods},
  pages={3--17},
  year={2007},
  publisher={Springer},
}

@article{mathiassen2002collaborative,
  title={Collaborative practice research},
  author={Mathiassen, Lars},
  journal={Information Technology \& People},
  volume={15},
  number={4},
  pages={321--345},
  year={2002},
  publisher={MCB UP Ltd},
}

@article{Mathiassenetal2012,
   author = {Mathiassen, Lars and Chiasson, Mike and Germonprez, Matt},
   title = {Style composition in action research publication},
   journal = {MIS quarterly},
   pages = {347-363},
   ISSN = {0276-7783},
   year = {2012},
   type = {Journal Article},
}

@book{OOAD2000,
   author = {Mathiassen, Lars and Munk-Madsen, Andreas and Nielsen, Peter A and Stage, Jan},
   title = {Object-oriented analysis \& design},
   publisher = {Forlaget Marko},
   address = {Aalborg},
   volume = {25},
   year = {2000},
   type = {Book},
}

@mastersthesis{Nguyen,
   author = {Nguyen Hoang, Vuong  },
   title = {Improving e-learning in maritime education and training: action research in the Vietnam maritime context},
   university = {World Maritime University},
    school = {World Maritime University},
   year = {2021},
   type = {Thesis},
}

@book{nicolini2012practice,
  title={Practice theory, work, and organization: An introduction},
  author={Nicolini, Davide},
  year={2012},
  publisher={OUP Oxford},
}

@article{PipekWulf2009,
   author = {Pipek, Volkmar. and Wulf, Volker.},
   title = {Infrastructuring: Towards an Integrated Perspective on the Design and Use of Information Technology},
   journal = {Journal of the Association for Information Systems},
   volume = {10},
   number = {5},
   pages = {447-473},
   year = {2009},
   type = {Journal Article},
}

@book{RobsonMcCartan,
   author = {Robson, Colin and McCartan, Kieran},
   title = {Real world research: A resource for users of social research methods in applied settings},
   publisher = {Wiley},
   address = {Hoboken},
   year = {2016},
   type = {Book},
}

@phdthesis{Ronkko2005,
   author = {Rönkkö, Kari},
   title = {Making Methods Work in Software Engineering},
   university = {Blekinge Institute of Technology},
    school = {Blekinge Institute of Technology},
   year = {2005},
   type = {Thesis},
}

@inproceedings{ronkkoetal2008,
   author = {Rönkkö, K. and Hellman, M. and Dittrich, Y.},
   title = {PD method and socio-political context of the development organization},
   booktitle = {Proceedings of the Tenth Anniversary Conference on Participatory Design 2008},
   publisher = {Indiana University},
   year = {2008},
   pages = {71-80},
   ISBN = {0981856101},
   url = {http://dl.acm.org/citation.cfm?id=1795245},
   type = {Conference Proceedings},
}

@article{Ronkkoetal2004,
   author = {Rönkkö, Kari and Hellman, M. and Kilander, B. and Dittrich, Yvonne},
   title = {Personas is not applicable: local remedies interpreted in a wider context},
   journal = {Proceedings of the eighth conference on Participatory design: Artful integration: interweaving media, materials and practices-Volume 1},
   pages = {112-120},
   year = {2004},
   type = {Journal Article},
}

@phdthesis{Seidelin2020,
   author = {Seidelin, Cathrine},
   title = {Towards a co-design perspective on data: Foregrounding data in the design and innovation of data-based services},
   university = { IT University of Copenhagen},
    school = { IT University of Copenhagen},
   year = {2020},
   type = {Thesis}
}

@inproceedings{SeidelinDittrichGronvall2018data,
  title={Data Work in a Knowledge-Broker Organization: How Cross-Organizational Data Maintenance shapes Human Data Interactions.},
  author={Seidelin, Cathrine and Dittrich, Yvonne and Gr{\"o}nvall, Erik},
  booktitle={British HCI Conference 2018},
  pages={1--12},
  year={2018},
  organization={BCS Learning and Development Ltd}
}

@article{SeidelinDittrichGronval2020,
   author = {Seidelin, Cathrine and Dittrich, Yvonne and Grönvall, Erik},
   title = {Foregrounding data in co-design – An exploration of how data may become an object of design},
   journal = {International Journal of Human-Computer Studies},
   volume = {143},
   pages = {102505},
   ISSN = {1071-5819},
   year = {2020},
   type = {Journal Article}
}

@inproceedings{SeidelinDittrichGronvall2020co,
  title={Co-designing data experiments: Domain experts’ exploration and experimentation with self-selected data sources},
  author={Seidelin, Cathrine and Dittrich, Yvonne and Gr{\"o}nvall, Erik},
  booktitle={Proceedings of the 11th nordic conference on human-computer interaction: shaping experiences, shaping society},
  pages={1--11},
  year={2020}
}

@inproceedings{SeidelinLeeDittrich2020,
  title={Understanding data and cooperation in a public sector arena},
  author={Seidelin, Cathrine and Lee, Charlotte P and Dittrich, Yvonne},
  booktitle={Proceedings of 18th European Conference on Computer-Supported Cooperative Work},
  volume={4},
  number={1},
  year={2020},
  organization={European Society for Socially Embedded Technologies (EUSSET)}
}

@inproceedings{SeidelinSivertsenDittrich2020,
   author = "Cathrine Seidelin and {Moeslund Sivertsen}, Stine and Yvonne Dittrich",
    title = "Designing an organisation{\textquoteright}s design culture: How appropriation of service design tools and methods cultivates sustainable design capabilities in SMEs",

year = "2020",
language = "English",
isbn = "978-91-7929-779-4",
pages = "11--31",
editor = "Yoko Akama and Liam Fennessy and Sara Harrington and Anna Farago",
booktitle = "ServDes.2020 Tensions, Paradoxes and Plurality Conference Proceedings, 2-5th February 2021, Melbourne, Australia",
note = "ServDes2020 Tensions, Paradoxes and Plurality Conference Proceedings, 2-5th February 2021, Melbourne, Australia ; Conference date: 02-02-2021 Through 05-02-2021",
url = "https://servdes2020.org",
}

@inbook{Serrat2017,
   author = {Serrat, Olivier},
   title = {The Five Whys Technique},
   booktitle = {Knowledge Solutions: Tools, Methods, and Approaches to Drive Organizational Performance},
   editor = {Serrat, Olivier},
   publisher = {Springer Singapore},
   address = {Singapore},
   pages = {307-310},
   ISBN = {978-981-10-0983-9},
   DOI = {10.1007/978-981-10-0983-9-32},
   year = {2017},
   type = {Book Section}
}

@article{star1989institutional,
  title={Institutional ecology,translations' and boundary objects: Amateurs and professionals in Berkeley's Museum of Vertebrate Zoology, 1907-39},
  author={Star, Susan Leigh and Griesemer, James R},
  journal={Social studies of science},
  volume={19},
  number={3},
  pages={387--420},
  year={1989},
  publisher={Sage Publications London}
}

@phdthesis{Unphon2010a,
title = {Re-engineering for Evolvability: Considering social as well as technical requirements for software products},
author = {Hataichanok Unphon},
year = {2010},
publisher = {IT University of Copenhagen},
school = {IT University of Copenhagen},
address = {Denmark},

}

@unpublished{Unphon2010,
   author = {Unphon, Hataichanok},
   title = {Architecture-Level Evolvability Assessment: Assessing Sustainability of Software Product Evolution},
   publisher = {IT University of Copenhagen},
   year = {2010},
   type = {Unpublished Work}
}

@inproceedings{UphonDittrich2008,
title = "Organisation Matters: How the Organisation of Software Development Influences the Development of Product Line Architecture",
author = "Hataichanok Unphon and Yvonne Dittrich",
note = "er den peer-reviewed ? Ja.; Software Engineering, SE 2008, as part of the 26th IASTED International Multi-Conference on APPLIED INFORMATICS ; Conference date: 12-02-2008 Through 14-02-2008",
year = "2008",
language = "English",
isbn = "978-0-88986-715-4",
pages = "178--183",
booktitle = "Software Engineering, SE 2008, February 12 – 14, 2008 Innsbruck, Austria",
publisher = "ACTA Press",
address = "Canada",

}

@article{UnphonDittrich2010,
   author = {Unphon, Hataichanok and Dittrich, Yvonne},
   title = {Software architecture awareness in long-term software product evolution},
   journal = {Journal of Systems and Software},
   volume = {83},
   number = {11},
   pages = {2211-2226},
   ISSN = {0164-1212},
   year = {2010},
   type = {Journal Article}
}

@inproceedings{Unphon2009a,
title = "Making use of architecture throughout the software life cycle - How the build hierarchy can facilitate product line developments",
author = "Hataichanok Unphon",
year = "2009",
doi = "10.1109/SHARK.2009.5069114",
language = "English",
isbn = "978-1-4244-3726-9",
pages = "41--48",
booktitle = "Proceedings of the 2009 ICSE Workshop on Sharing and Reusing Architectural Knowledge",
publisher = "IEEE Computer Society Press",
address = "United States",
note = "International Conference on Software Engineering ; Conference date: 02-07-2010",

}
\end{document}